\documentclass{article}
\usepackage{spconf,amsmath,graphicx}
\usepackage{xcolor}
\usepackage{tablefootnote}
\usepackage[hidelinks]{hyperref}
\usepackage[noadjust]{cite}

\title{\vspace{-16mm}Upsampling artifacts in neural audio synthesis\vspace{-1mm}}
\name{Jordi Pons, Santiago Pascual, Giulio Cengarle, Joan Serr\`{a}\vspace{-3mm}}
\address{Dolby Laboratories\vspace{-3mm}}
\begin{document}
\ninept

\maketitle

\begin{abstract}

\noindent A number of recent advances in neural audio synthesis rely on upsampling layers, which can introduce undesired artifacts. 
In computer vision, upsampling artifacts have been studied and are known as checkerboard artifacts (due to their characteristic visual pattern). However, their effect has been overlooked so far in audio processing.
Here, we address this gap by studying this problem from the audio signal processing perspective. We first show that the main sources of upsampling artifacts are: 
(i)~the tonal and filtering artifacts introduced by problematic upsampling operators, 
and (ii) the spectral replicas that emerge while upsampling.
We then compare different upsampling layers, showing that nearest neighbor upsamplers can be an alternative to the problematic (but state-of-the-art) transposed and subpixel convolutions which are prone to introduce tonal artifacts.

\end{abstract}
\vspace{1.2mm}

\vspace{-2mm}
\section{introduction}
\vspace{-2mm}

Feed-forward neural audio synthesizers~\cite{donahue2018adversarial, stoller2018wave,defossez2019music,kumar2019melgan} were recently proposed as an alternative to Wavenet~\cite{oord2016wavenet}, which is 
computationally demanding and slow \mbox{due to its dense and auto-regressive nature~\cite{rethage2018wavenet}.}
Among the different feed-forward architectures proposed for neural audio synthesis~\cite{rethage2018wavenet,prenger2019waveglow,serra2019blow}, generative adversarial networks (GANs)~\cite{donahue2018adversarial,kumar2019melgan,pascual2017segan} and autoencoders~\cite{stoller2018wave,defossez2019music,pandey2020densely} heavily rely on upsampling layers.
GANs allow for efficient feed-forward generation, by upsampling low-dimentional vectors to waveforms.
Autoencoders also allow for feed-forward generation, and their bottleneck layers are downsampled---requiring less computational and memory footprint around the bottleneck.
While GANs and autoencoders allow for fast feed-forward architectures, the upsampling layers that are typically embedded in these models can introduce upsampling artifacts~\cite{donahue2018adversarial,stoller2018wave}.

\begin{figure}[h]
	\vspace{-1mm}
	\begin{minipage}[b]{0.49\linewidth}
		\centering
		\centerline{\includegraphics[width=0.78\linewidth]{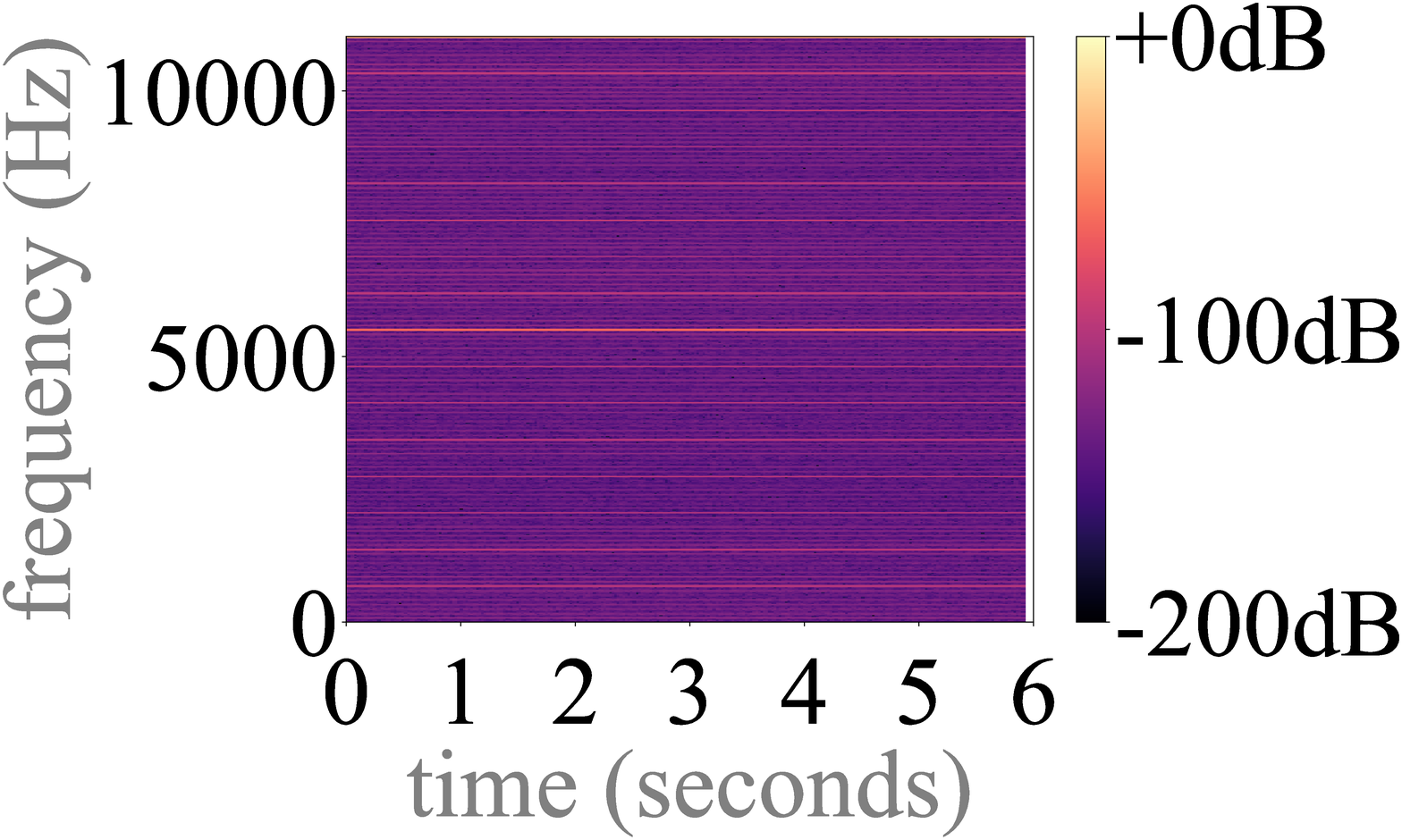}}
		\vspace{-0.32cm}
		{(a) MelGAN~\cite{kumar2019melgan}}\medskip
		\vspace{0.15cm}
	\end{minipage}
	\begin{minipage}[b]{0.49\linewidth}
		\centering
		\centerline{\includegraphics[width=0.78\linewidth]{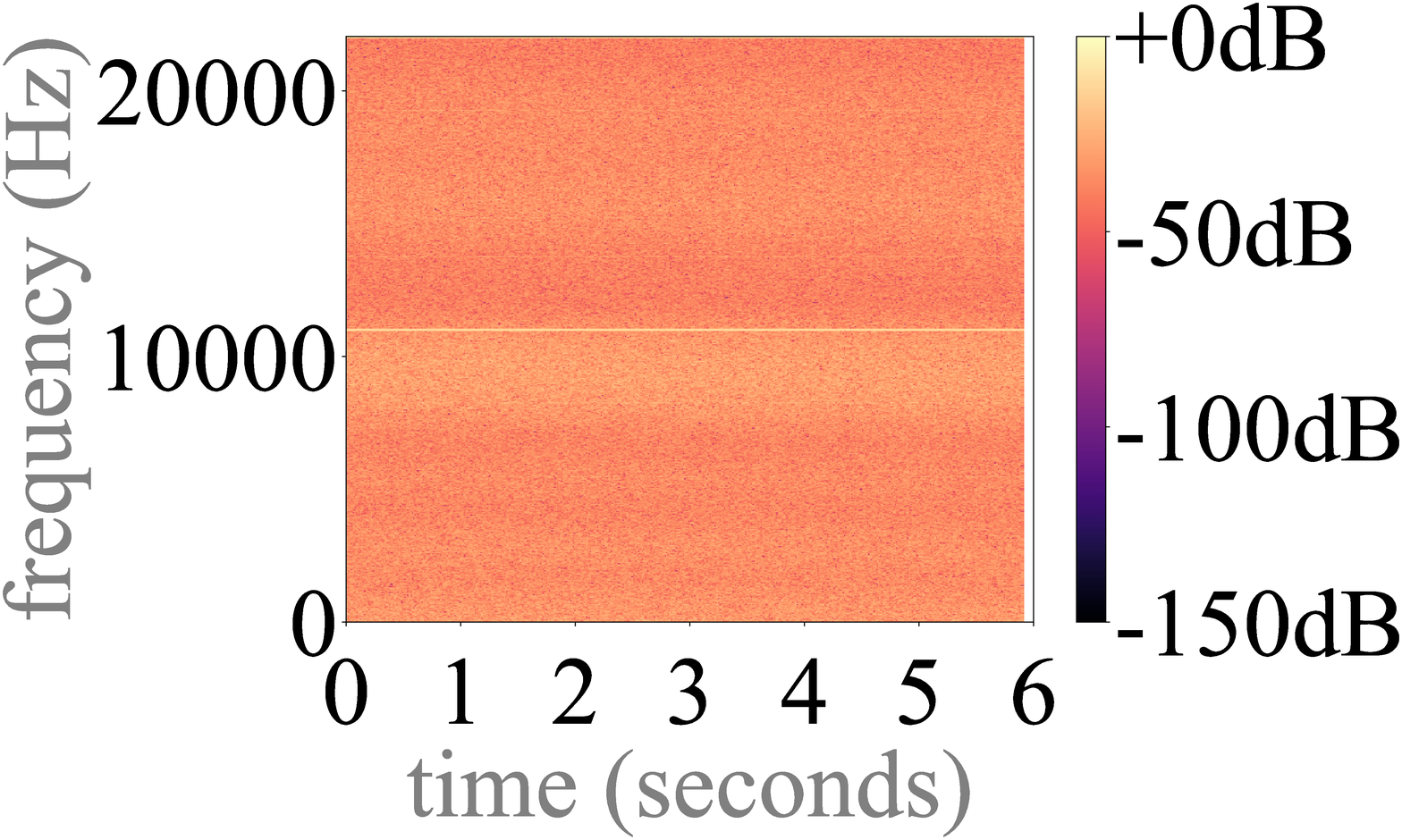}}
		\vspace{-0.32cm}
		{(b) Demucs~\cite{defossez2019music}: original}\medskip
		\vspace{0.15cm}
	\end{minipage}
	\hfill		\vspace{-0.5cm}
	\begin{minipage}[b]{.49\linewidth}
		\centering
		\centerline{\includegraphics[width=0.78\linewidth]{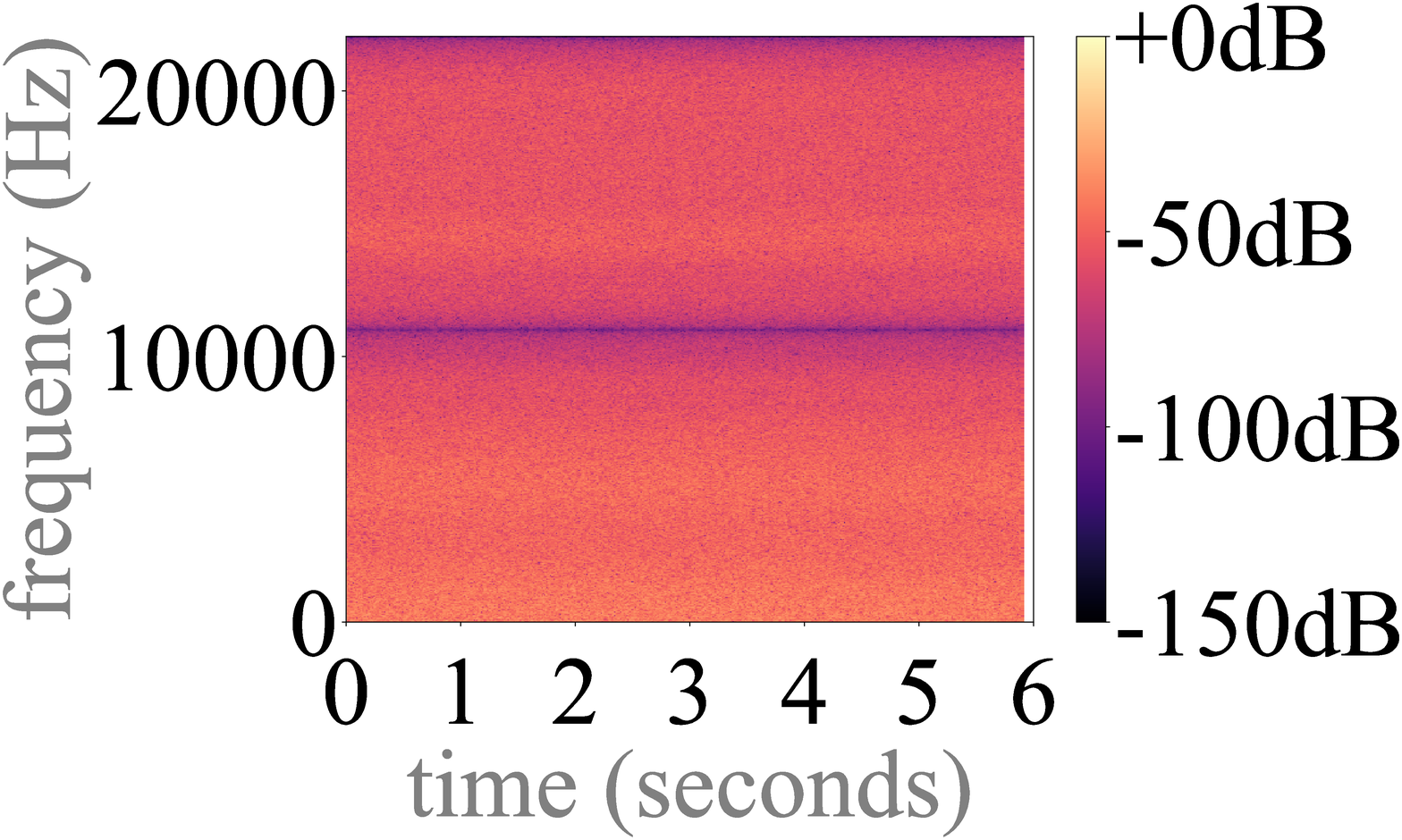}}
		\vspace{-0.32cm}
		\mbox{(c) Demucs: nearest neighbor}\medskip
		\vspace{0.15cm}
	\end{minipage}
	\begin{minipage}[b]{0.49\linewidth}
		\centering
		\centerline{\includegraphics[width=0.78\linewidth]{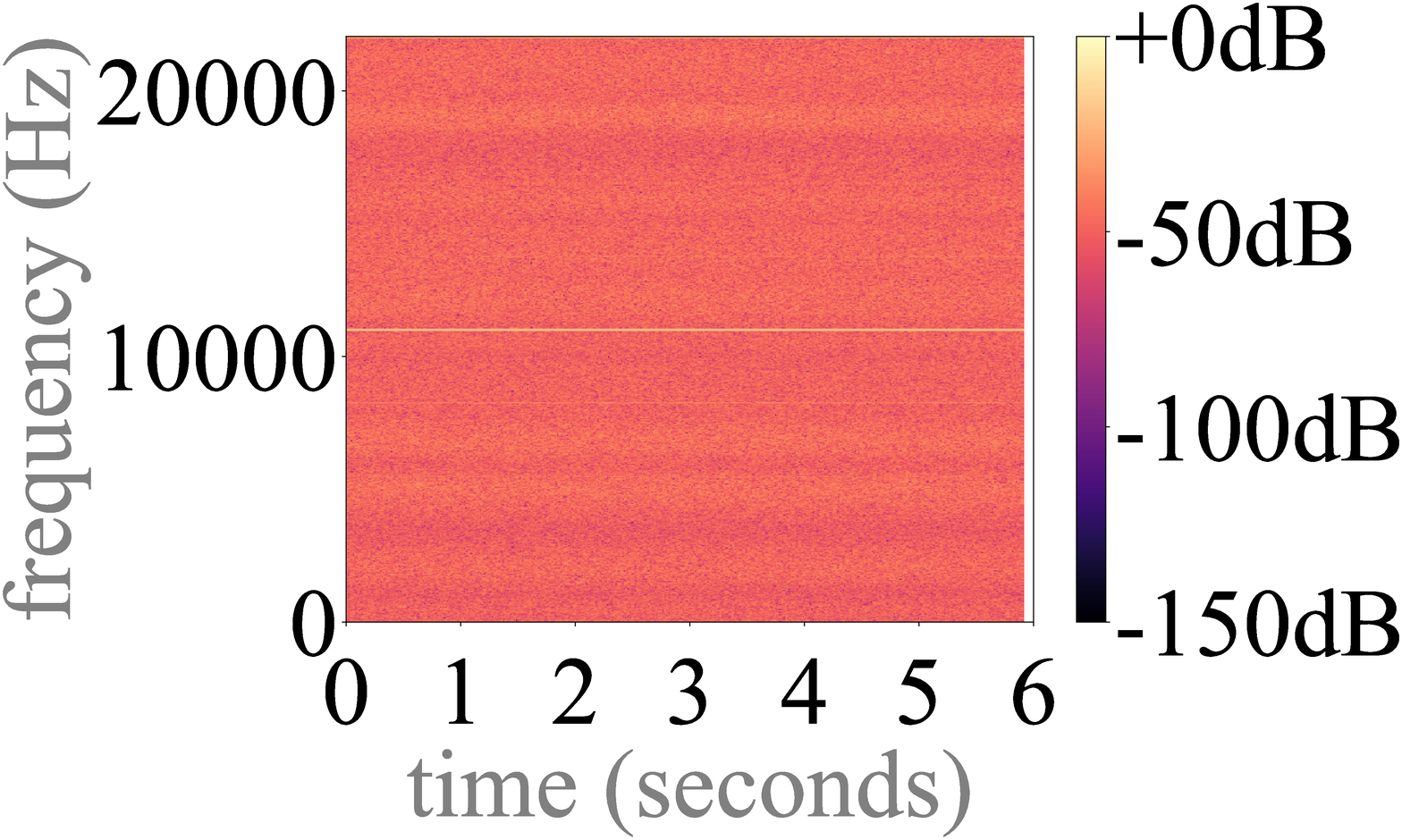}}
		\vspace{-0.32cm}
		{(d) Demucs: subpixel CNN}\medskip
		\vspace{0.15cm}
	\end{minipage}
	\hfill	
	
	\vspace{-6mm}
	\caption{\textbf{Upsampling artifacts after initialization}: tonal artifacts (horizontal lines: a,b,d) and {filtering artifacts (horizontal valley: c).} 
		 Input: white noise. {MelGAN operates at 22kHz, Demucs at 44kHz.}}
	\label{fig:all}
		\vspace{-5mm}
\end{figure}

Three main types of upsampling layers exist:
transposed convolutions~\cite{donahue2018adversarial,defossez2019music,pascual2017segan}, interpolation upsamplers~\cite{stoller2018wave,binkowski2019high}, and {subpixel convolutions}~\cite{pandey2020densely,kuleshov2017audio,chou2018multi}.
These can introduce upsampling artifacts, as shown in Fig.~\ref{fig:all}---where we plot a spectrogram of their output after random initialization,
to stress that upsampling artifacts are already present before training.
We experiment with state-of-the-art neural synthesizers based on transposed convolutions (MelGAN and Demucs autoencoder) or on alternative upsampling layers (interpolation and subpixel convolutions).
MelGAN uses transposed convolutions for upsampling~\cite{kumar2019melgan}. We implement it as {Kumar \textit{et al.}~\cite{kumar2019melgan}}: with 4 transposed convolution layers of \textit{length=16,16,4,4} and \textit{stride=8,8,2,2} respectively. 
Demucs is an autoencoder employing transposed convolutions. We implement it as \mbox{D{\'e}fossez \textit{et al.}~\cite{defossez2019music}:} with 6  transposed convolution layers of \textit{length=8} and \textit{stride=4}. 
Transposed convolutions in MelGAN and Demucs introduce what we call ``tonal artifacts" after initialization (Fig.~\ref{fig:all}:~a,~b, and Sec.~2).
Next, we modify Demucs' upsampling layers to rely on nearest neigbor interpolation~\cite{gritsenko2020spectral} or subpixel convolution~\cite{kuleshov2017audio} upsamplers. Interpolation upsamplers can introduce  what we describe as ``filtering artifacts" {(Fig.~\ref{fig:all}: c, and Sec.~3)}, while subpixel convolution can also introduce the above mentioned ``tonal artifacts" \mbox{(Fig.~\ref{fig:all}: d, and Sec.~4)}.
{In sections 2, 3 and 4, we describe the origin of these upsampling artifacts. In section 5, we note that spectral replicas can introduce additional artifacts. Finally, in section~6, we
 discuss the effect that training can have on such artifacts.} A notebook \mbox{version of the paper, with code to experiment with the figures below,} is at: \href{https://github.com/DolbyLaboratories/neural-upsampling-artifacts-audio}{\textit{github@dolbylaboratories/neural-upsampling-artifacts-audio}}.

\section{ transposed convolutions}
	\vspace{-2mm}
Transposed CNNs are widely used for audio synthesis~\cite{donahue2018adversarial, defossez2019music,pascual2017segan} and can introduce tonal {artifacts due to~\cite{odena2016deconvolution}: (i)~their {weights' initialization},} {(ii)~{overlap} issues, and (iii)~the {loss function}}.
Issues \mbox{(i) and (ii)} are related to the model's initialization \mbox{and construction, respectively,} while issue (iii)~depends on how learning is defined.~In this article, we use the terms \textit{length} and \textit{stride} for referring to the transposed~convolution filter length and stride, respectively.
Three situations arise:

\noindent \hspace{2mm}-- \underline{No overlap}: \textit{length=stride}. No overlap artifacts are introduced, but the weight initialization issue can introduce tonal artifacts.

\noindent \hspace{2mm}-- \underline{Partial overlap}: {\textit{length} is not a multiple of \textit{stride}.} Overlap and weight initialization issues can \mbox{introduce tonal and boundary artifacts.}

\noindent \hspace{2mm}-- \underline{Full overlap}: \textit{length} is a multiple of \textit{stride}. Overlap artifacts can be introduced (as boundary artifacts at the borders) and the weight initialization issue introduces tonal artifacts after initialization.

\vspace{0.4mm} 

\begin{figure}[ht]
	\vspace{-1mm}
	\centering
	\centerline{\includegraphics[width=4.4cm]{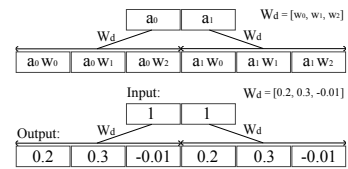}}
	\vspace{-1.5mm}
	
	\caption{\textbf{Transposed convolution}: \textit{length=stride=3}, w/o bias. The example depicts a periodicity every 3 samples, at the stride length.}
	\label{fig:1}
	\vspace{-2mm}
\end{figure}

\noindent \textbf{Issue (i): Weight Initialization} --- It is caused by the transposed convolution weights that repeat across time, generating a periodic pattern (tonal artifacts) after random initialization. To understand this issue, we leave the overlap and loss function issues aside and consider an example of a randomly initialized transposed convolution with {no overlap} (\mbox{\textit{length=stride=3}}, see Fig.~\ref{fig:1}). 
Given that the filter $W_d$ is shared across time, 
the resulting feature map includes a periodicity related to the temporal structure of the (randomly initialized) weights. 
Fig.~\ref{fig:1} (bottom) exemplifies this behavior by feeding ones to the above-mentioned transposed convolution:
the temporal patterns present in the (random) weights introduce high-frequency periodicities that call to be compensated by training.
Note that with \textit{stride=length}, the emerging periodicities are dominated by the stride and length parameter.
Solutions to this issue involve using constant weights or alternative upsamplers.
{Sections~3 and 4~focus} on describing alternative upsamplers, since using constant weights can affect {expressiveness---and learning due to a poor (constant) initialization. }

\begin{figure}[ht]
	\vspace{-2mm}
	\centering
	\centerline{\includegraphics[width=0.74\linewidth]{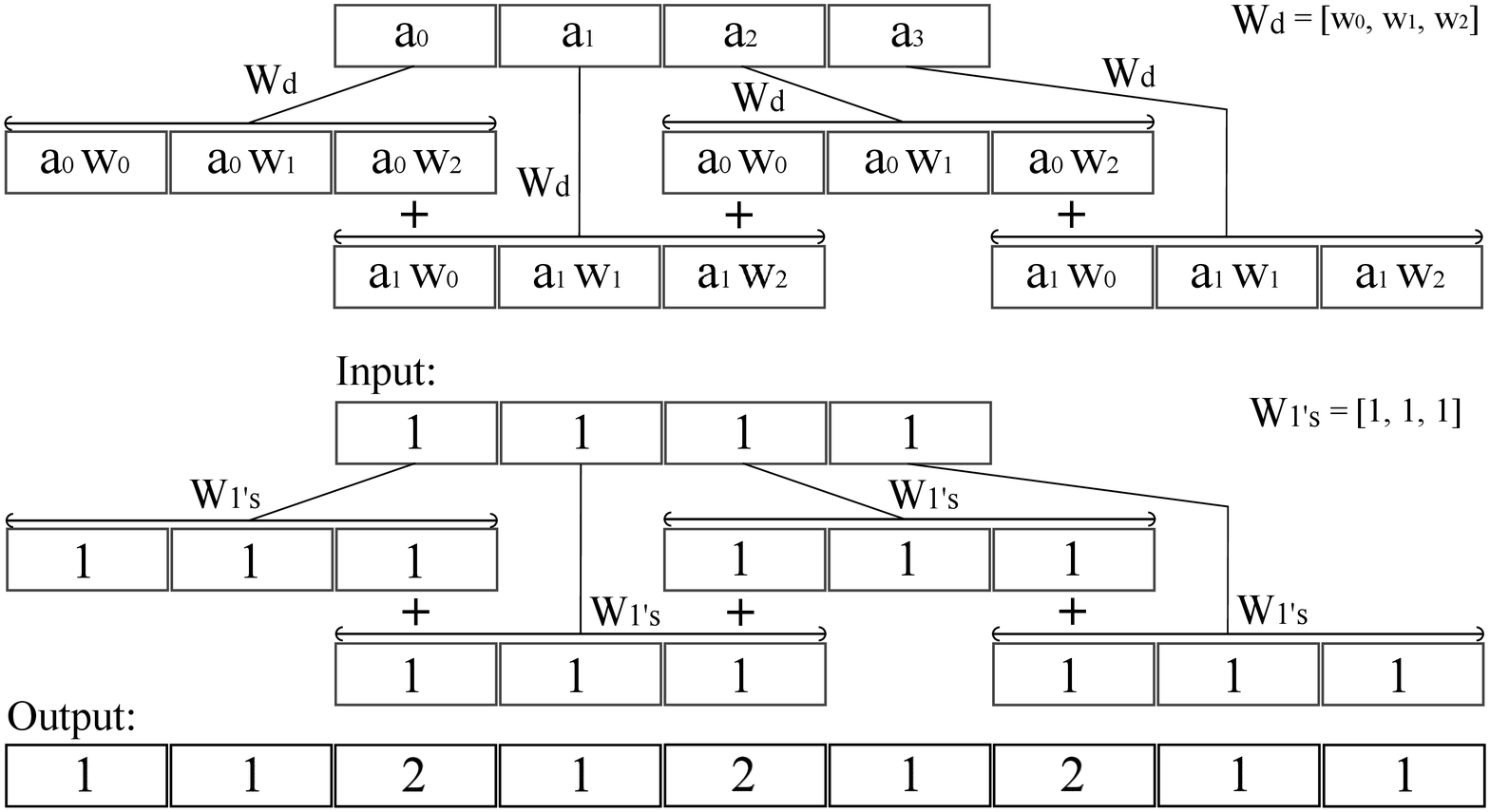}}
	\vspace{-0.2cm}
	\caption{\textbf{Transposed convolution}: \textit{length=3, stride=2}, w/o bias. The example depicts a periodicity every 2 samples, at the stride length.}
	\label{fig:2}
	\vspace{-2mm}
\end{figure}

\begin{figure}[t]
	\vspace{-1mm}
	\centering
	\centerline{\includegraphics[width=4.3cm]{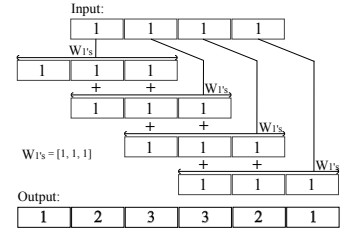}}
	\vspace{-0.2cm}
	\caption{\textbf{Transposed convolution}: \textit{length=3, stride=1}, w/o bias. Note: (i) artifacts due to no overlap at the borders, (ii) no periodicities emerge due to full overlap, (iii) it {barely upsamples (\textit{stride=1}).}}
	\label{fig:3}
	\vspace{-3mm}
\end{figure}

\noindent\textbf{Issue (ii): Overlap} --- It arises because, when~{\textit{stride}$<$\textit{length}}, the upsampled representations overlap and add. To understand this issue, we leave the weight initialization and loss function issues aside and consider constantly-initialized transposed convolutions with overlap (as in Figs.~\ref{fig:2} and~\ref{fig:3}).
Note that the \textit{stride}$>$\textit{length} setup is generally not desirable because parts of the upsampled signal are not defined. 
Due to the \mbox{\textit{stride}$<$\textit{length}} setup, upsampled representations overlap and add (see Figs.~\ref{fig:2} and~\ref{fig:3}). This can cause different problems, depending on the situation: {partial or full overlap}.

\noindent \hspace{2mm}-- \underline{Partial overlap}: \textit{length} is not a multiple of \textit{stride}. Upsampled representations are summed up (not averaged) and introduce a high-frequency periodicity to the signal. This behavior becomes clear if ones are fed to a \textit{length=3, stride=2} transposed convolution filter constantly
initialized with ones (Fig.~\ref{fig:2}, bottom).
A high-frequency pattern (tonal noise) emerges because the amount of overlap varies across the upsampled signal due to partial overlap. In addition, boundary artifacts \mbox{emerge due to {non-constant overlap} at the borders.}

\begin{figure}[ht]
	\vspace{-0mm}
	\centering
	\hspace{-9mm}\centerline{\includegraphics[width=0.89\linewidth]{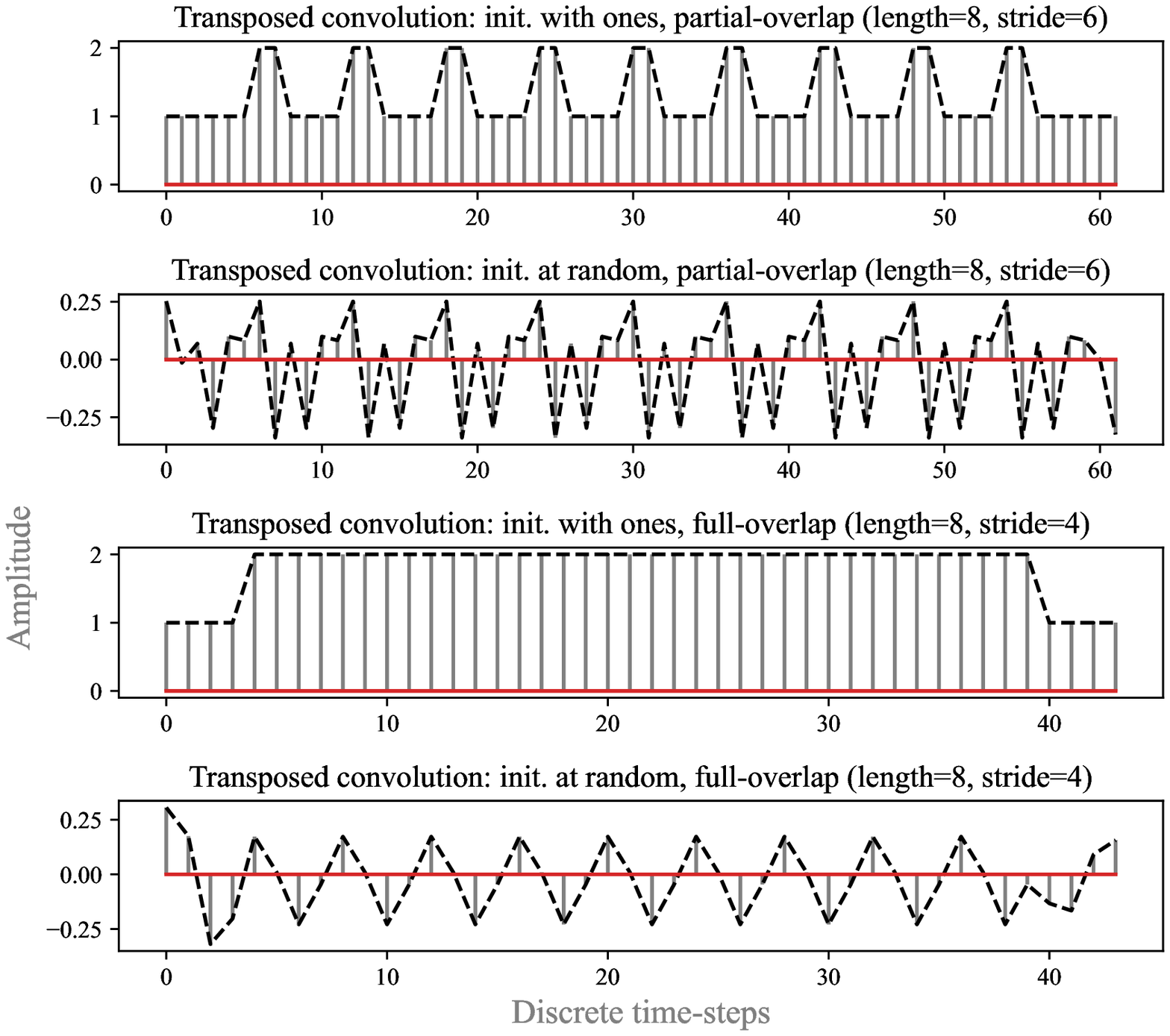}}
	
	\vspace{-6mm}
	\caption{\mbox{\textbf{Transposed convolutions}, w/o bias, with input set as ones.}}
	\label{fig:4}
	\vspace{-5mm}
\end{figure}

\noindent \hspace{2mm}-- \underline{Full overlap}: \textit{length} is a multiple of \textit{stride}. {Overlap-related tonal artifacts are not present because the amount of overlap is constant across the upsampled signal. }This behavior becomes clear if ones are fed to a \textit{length=3, stride=1} transposed convolution filter constantly initialized with ones (Fig.~\ref{fig:3}).
While full overlap setups do not introduce overlap-related tonal artifacts, boundary artifacts emerge due to {non-constant overlap} at the borders.
Fig.~\ref{fig:3} depicts a {full overlap} setup (with \textit{stride=1}) that {barely upsamples the signal}---but more effective full overlap configurations exist (see Fig.~\ref{fig:4}, row 3).

\vspace{0.6mm}
\noindent Fig.~\ref{fig:4} (row 1) shows that high-frequencies emerge due to partial overlap, which can be avoided using a full overlap setup \mbox{(row 3). }
However, as noted above, constant (ones) transposed convolution filters are rarely used.
Fig.~\ref{fig:4}~(rows 2, 4) illustrates the effect of using non-constant filters. Note that even when using a full overlap setup, a periodic pattern appears at the frequency of the stride. 
{Hence, importantly, the weight initialization issue remains.} Also note that boundary artifacts are introduced in both cases: for partial  and full overlap setups. 
Stoller \textit{et al.}~\cite{stoller2018wave} also described boundary artifacts.

\vspace{0.5mm}
\noindent \textbf{Issue (iii): Loss Function} --- Loss-related tonal artifacts can appear when using adversarial losses~\cite{donahue2018adversarial} and/or deep feature losses~\cite{germain2019speech}, since using CNNs as a loss involves a transposed convolution in the  backpropagation step that can cause high-frequency patterns in the gradient~\cite{odena2016deconvolution}. 
For example, loss-related high-frequency artifacts are noticeable when visualizing the learnt features of a neural network via optimization, since strided convolutions and pooling operations create high-frequency patterns in the gradient that  impact the resulting visualizations~\cite{olah2017feature}. 
Solutions proposed in the feature visualization literature involve relying on learnt priors~\cite{nguyen2016synthesizing} and/or regularizers~\cite{mahendran2015understanding,mordvintsev2015inceptionism}. 
Alternative solutions may consist in not using losses that involve a backpropagation step that is a transposed convolution.

\vspace{0.6mm}
\noindent In the context of adversarial training, Donahue \textit{et al.}~\cite{donahue2018adversarial} argued that high-frequency patterns are less common in images than in audio. Note that tonal artifacts may be similar to the (high) frequencies already present in real audio, which makes the discriminator's objective more challenging for audio~\cite{donahue2018adversarial}. Accordingly, adversarial losses might be more effective at removing transposed convolution artifacts for images than for~audio~\cite{donahue2018adversarial}. In addition, they also note that transposed convolution artifacts (tonal noise) will have a specific phase, allowing the discriminator to learn a trivial policy to detect {fake} audio. 
As a solution, they randomly {perturb the phase of the discriminator's feature maps~\cite{donahue2018adversarial}.}~

\vspace{0.4mm}
\noindent Finally,~note that the loss function issue remains understudied for audio (but also for computer vision~\cite{odena2016deconvolution,aitken2017checkerboard}). 
Hence, the impact of training models with such {losses has yet to be fully described.}

\begin{figure}[t]
	\vspace{-5mm}
	\begin{minipage}[b]{.49\linewidth}
		\centering
		\centerline{\includegraphics[width=0.78\linewidth]{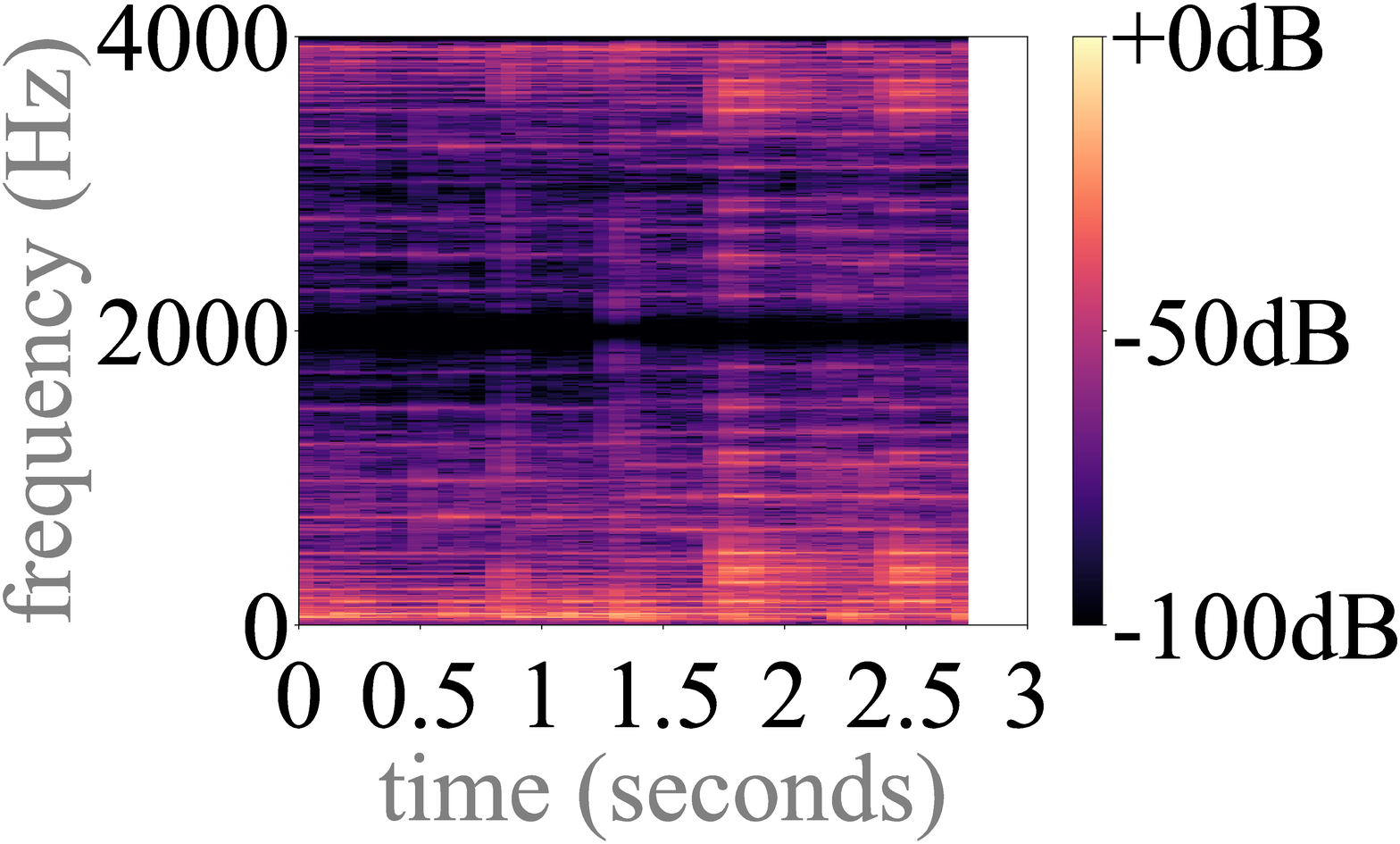}}
		\vspace{-0.3cm}
		{(a) Nearest neighbor: layer 1}\medskip
		\vspace{0.15cm}
	\end{minipage}
	\hfill		\vspace{-0.5cm}
	\begin{minipage}[b]{0.49\linewidth}
		\centering
		\centerline{\includegraphics[width=0.78\linewidth]{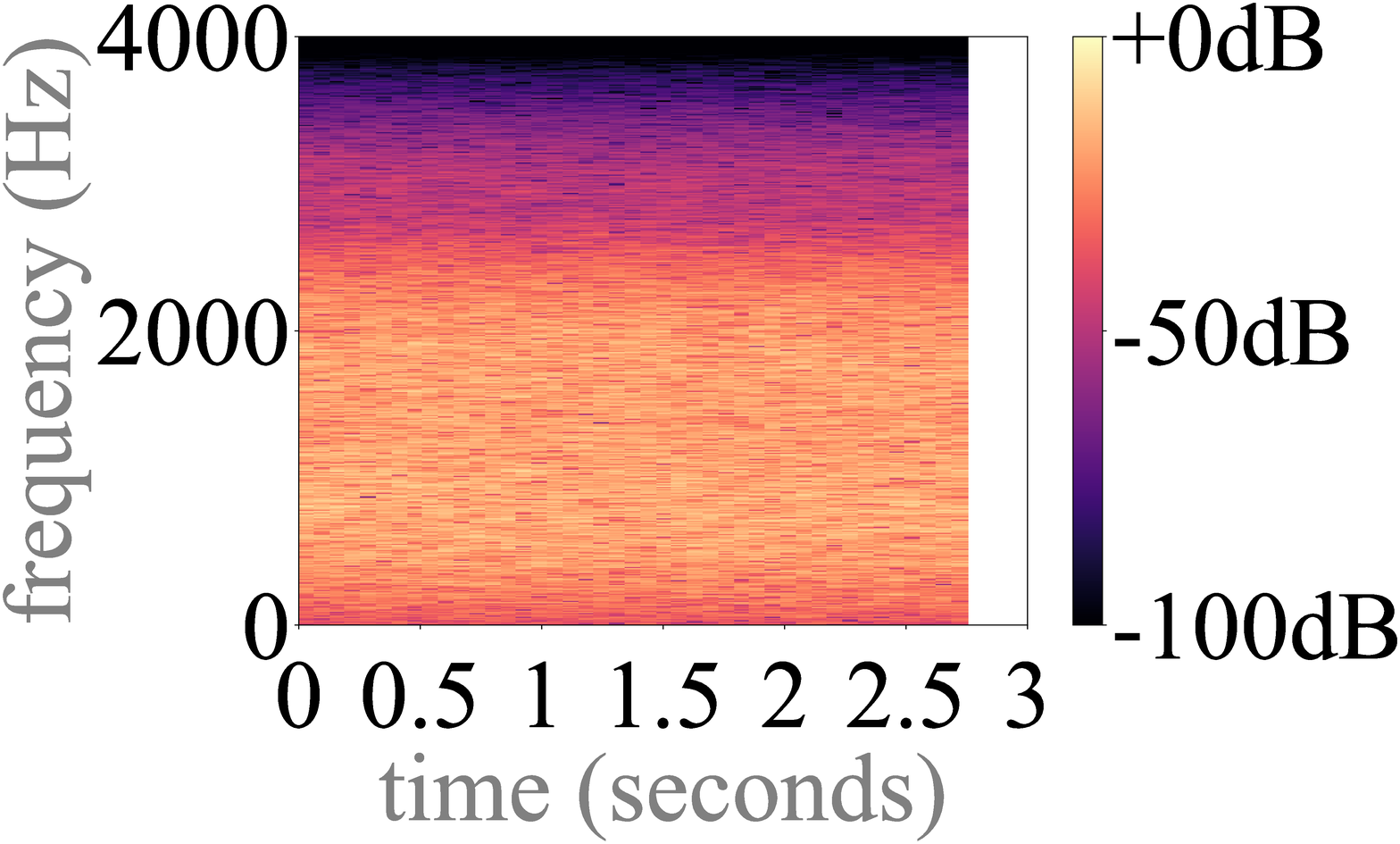}}
		\vspace{-0.3cm}
		{(d) Linear: layer 1}\medskip
		\vspace{0.15cm}
	\end{minipage}
	
	\begin{minipage}[b]{.49\linewidth}
		\centering
		\centerline{\includegraphics[width=0.78\linewidth]{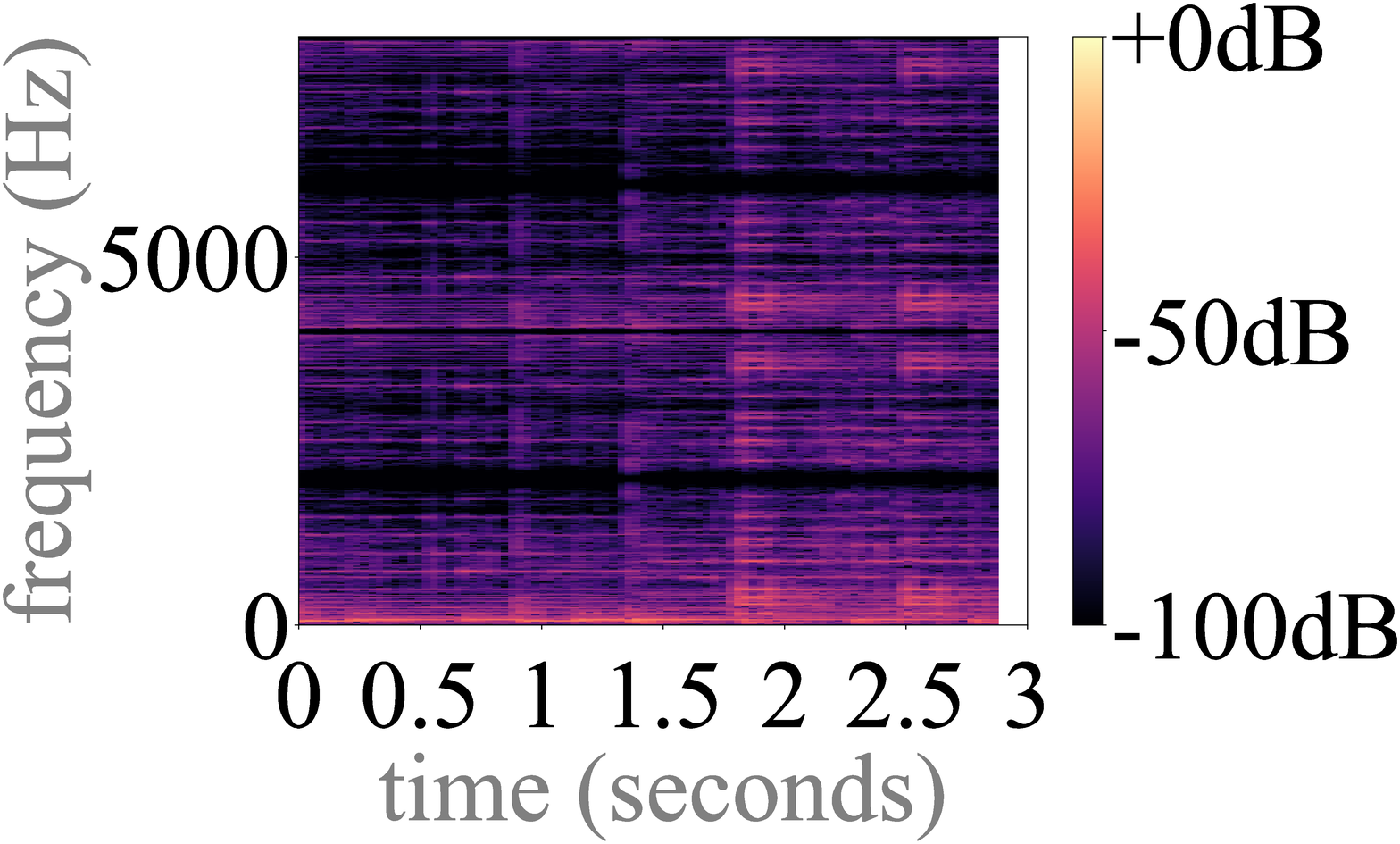}}
		\vspace{-0.3cm}
		{(b) Nearest neighbor: layer 2}\medskip
	\end{minipage}
	\hfill 	\vspace{-0.4cm}
	\begin{minipage}[b]{0.49\linewidth}
		\centering
		\centerline{\includegraphics[width=0.78\linewidth]{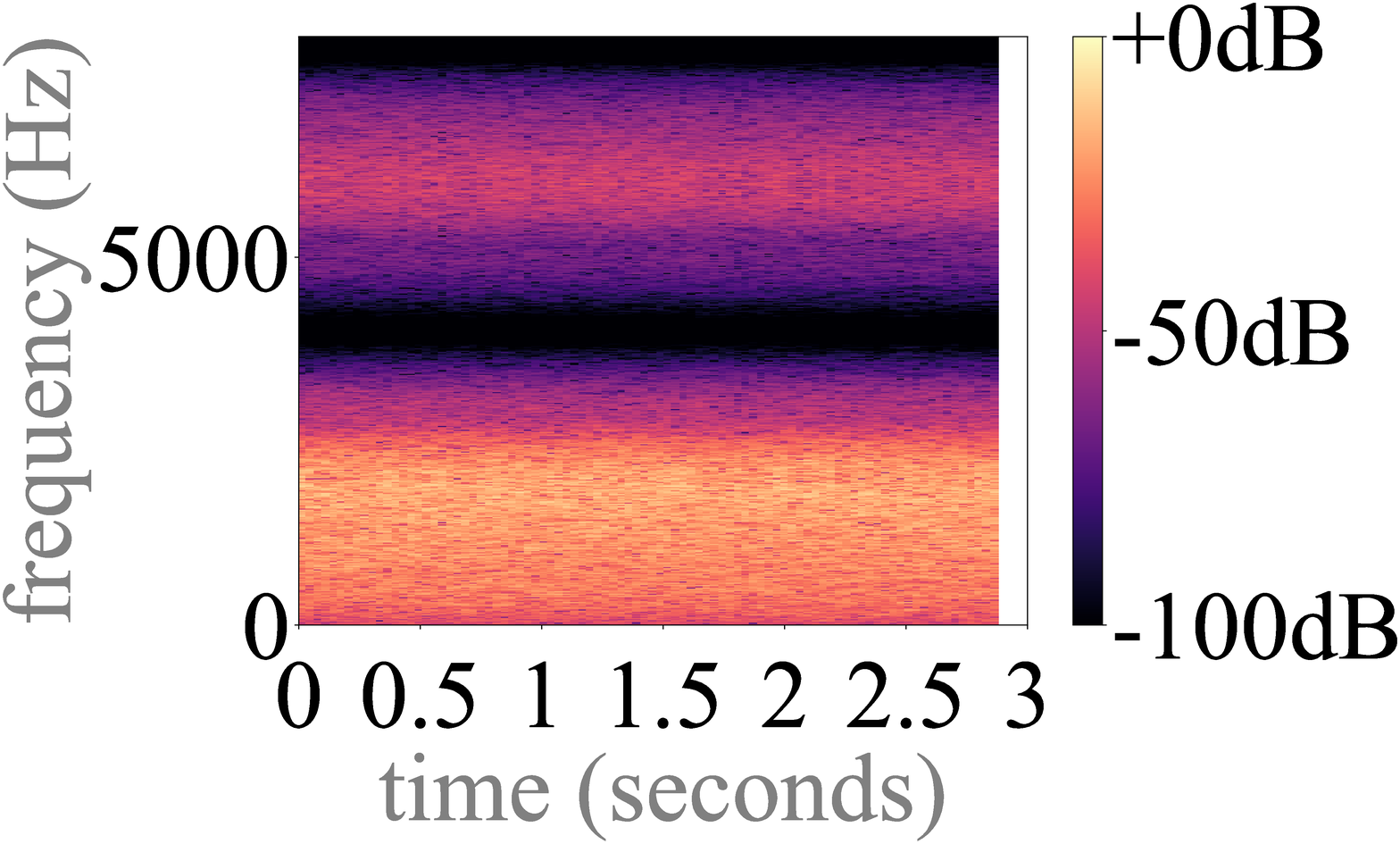}}
		\vspace{-0.3cm}
		{(e) Linear: layer 2}\medskip 
	\end{minipage}
	\begin{minipage}[b]{.49\linewidth}
		\centering
		\centerline{\includegraphics[width=0.78\linewidth]{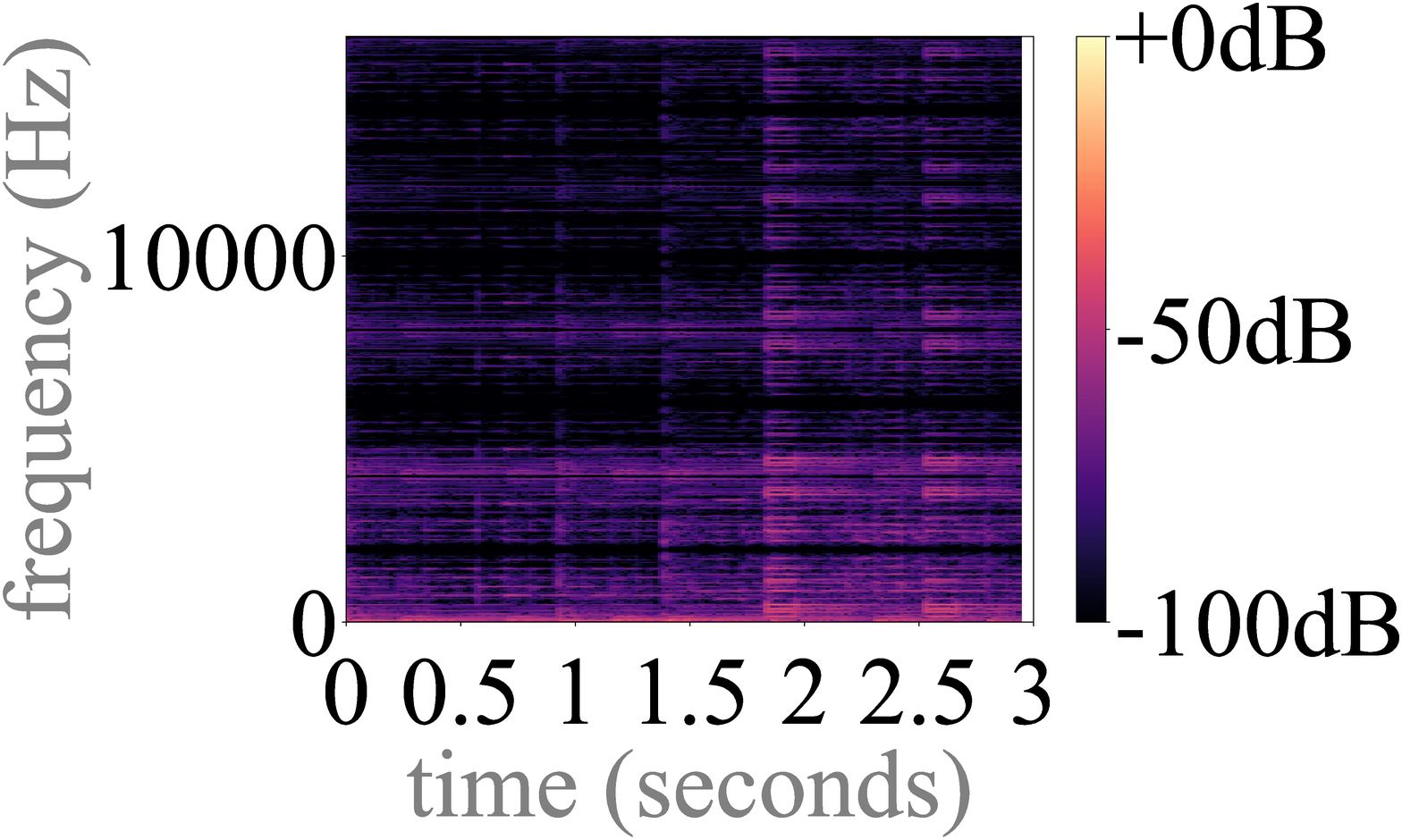}}
		\vspace{-0.3cm}
		{(c) Nearest neighbor: layer 3}\medskip
	\end{minipage}
	\hfill
	\begin{minipage}[b]{0.49\linewidth}
		\centering
		\centerline{\includegraphics[width=0.78\linewidth]{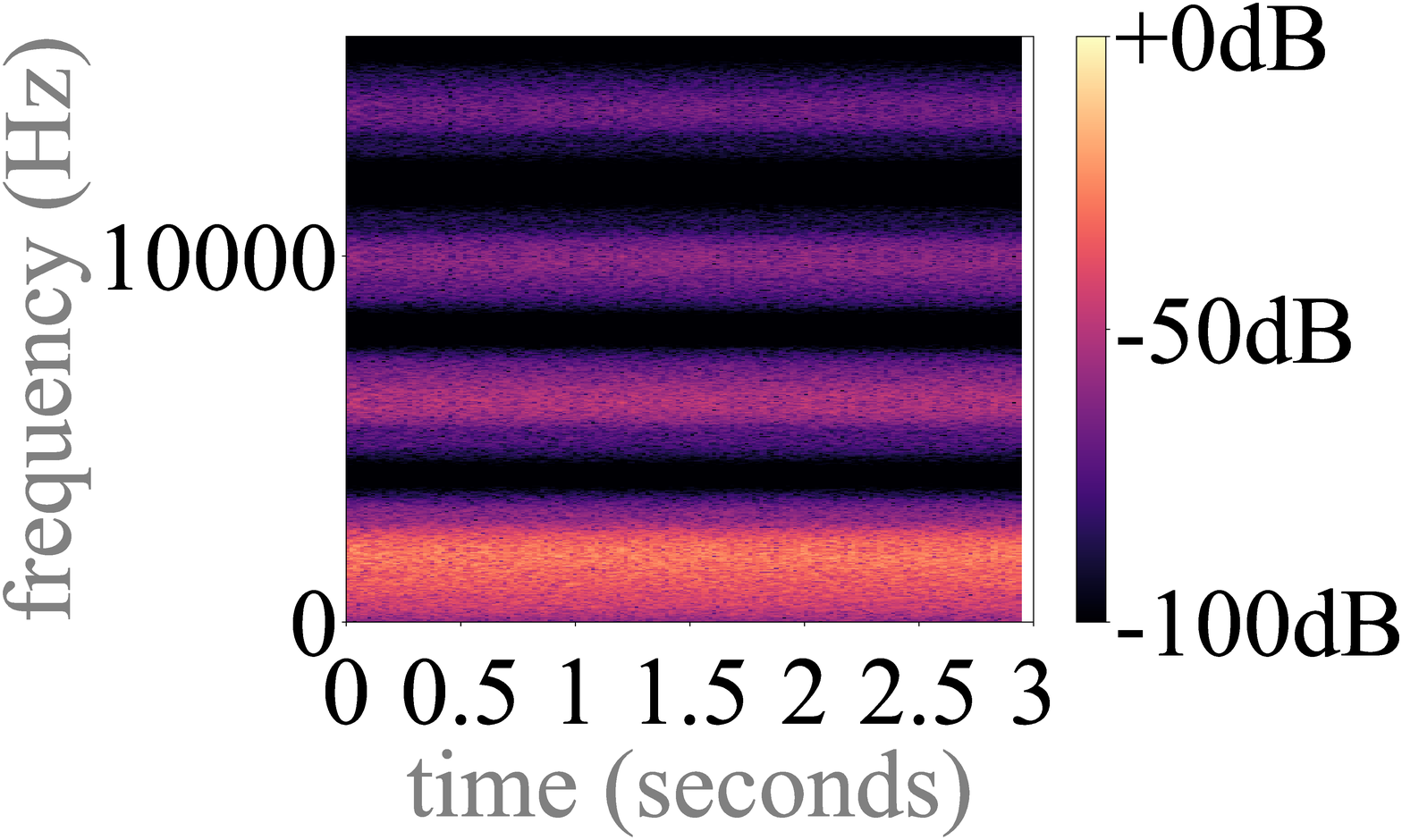}}
		\vspace{-0.3cm}
		{(f) Linear: layer 3}\medskip 
	\end{minipage}
	\vspace{-5mm}
	\caption{{\textbf{Interpolation upsamplers}: filtering artifacts, but no tonal artifacts, after initialization.~Each consecutive layer (top to bottom): nearest neighbor or linear interpolation (x2) + CNN (filters of length 9, stride 1). Inputs at 4kHz: music (left), white noise (right).}}
	\label{fig:interpolation}
	\vspace{-3mm}
\end{figure}

\begin{figure}[t]
	\vspace{-2mm}
	\begin{minipage}[b]{.49\linewidth}
		\centering
		\centerline{\includegraphics[width=0.78\linewidth]{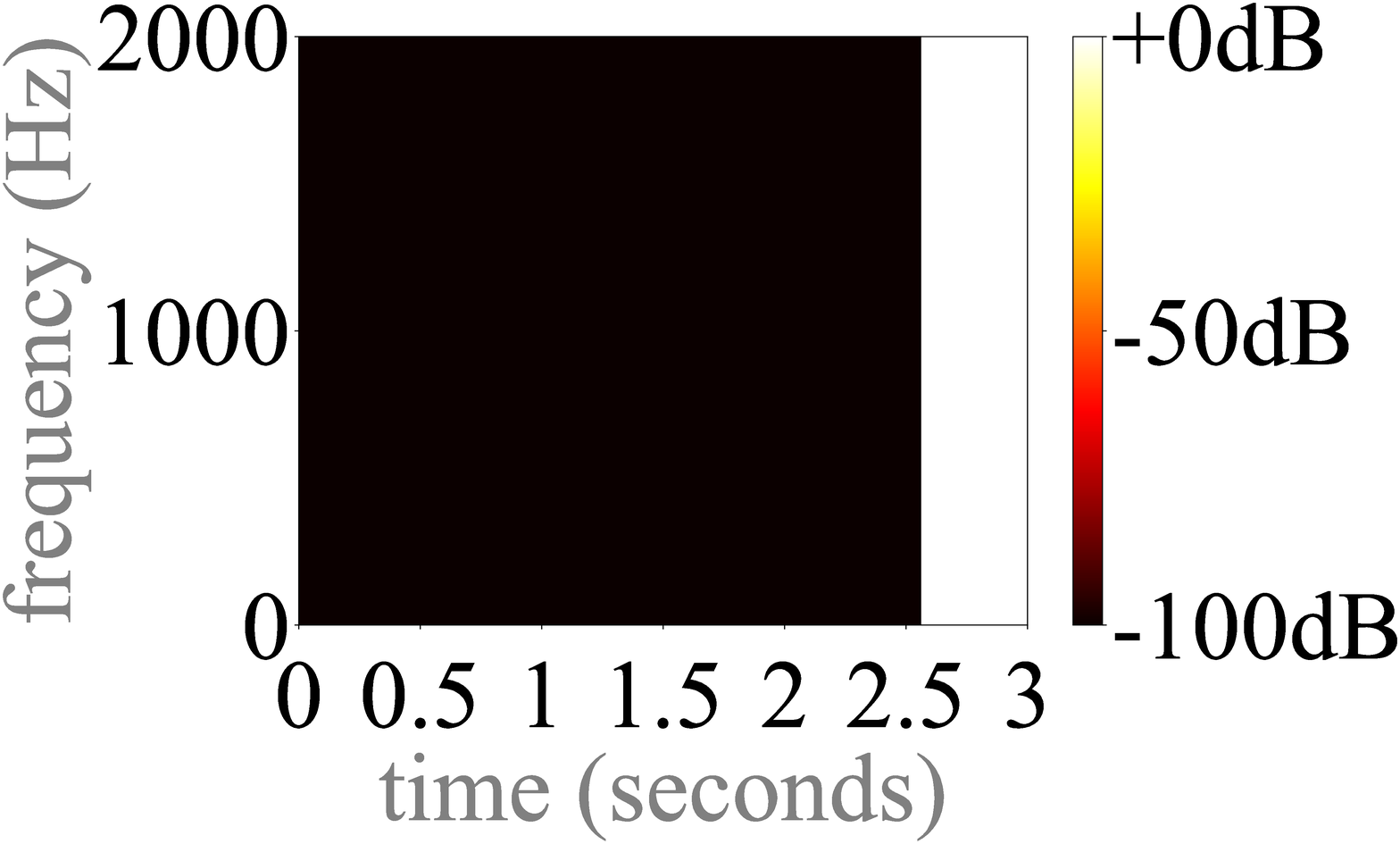}}
		\vspace{-0.3cm}
		{(a) Input: ones at 4kHz}\medskip
		\vspace{0.15cm}
	\end{minipage}
	\hfill		\vspace{-0.5cm}
	\begin{minipage}[b]{0.49\linewidth}
		\centering
		\centerline{\includegraphics[width=0.78\linewidth]{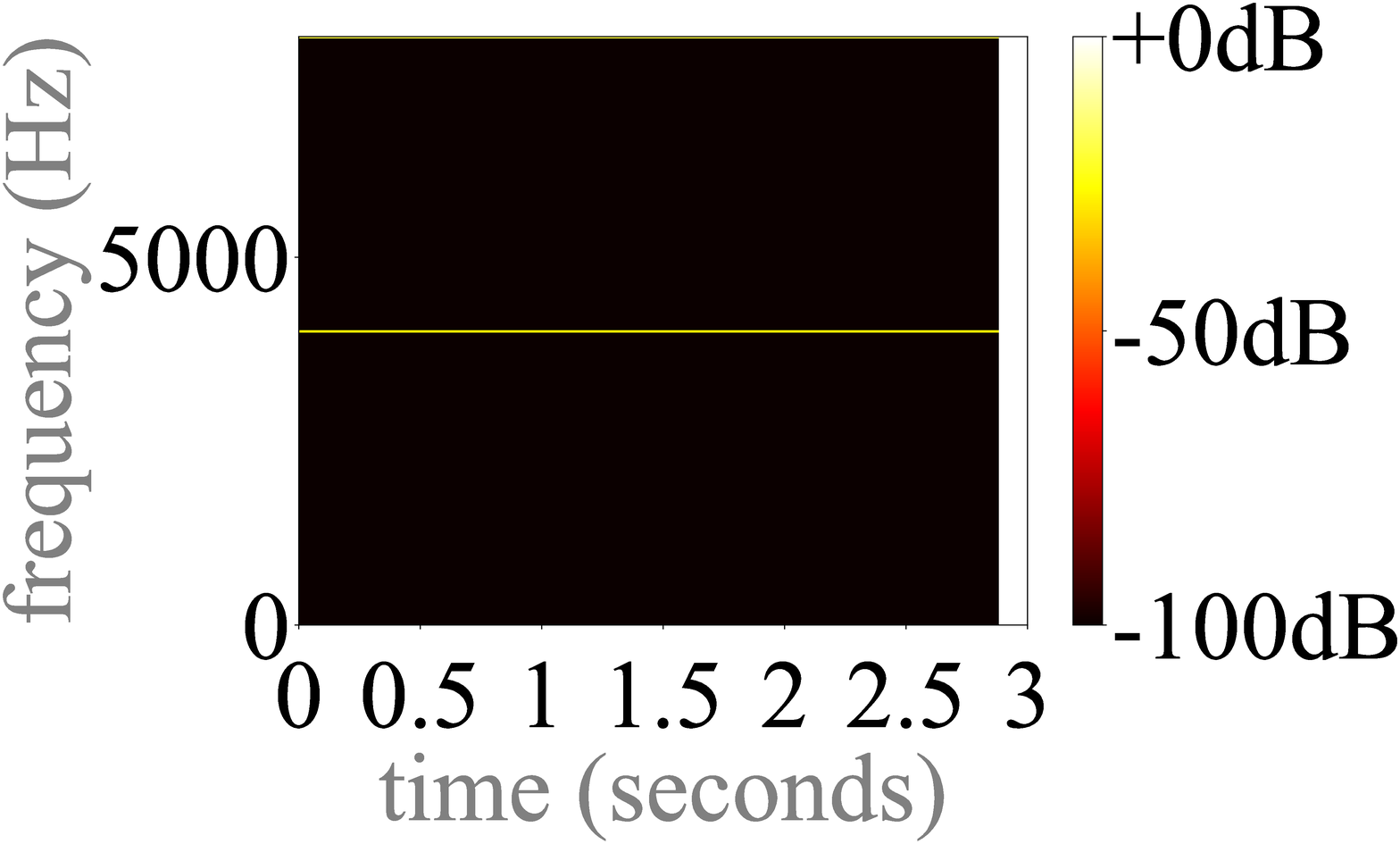}}
		\vspace{-0.3cm}
		{(d) subpixel CNN: layer 2}\medskip
		\vspace{0.15cm}
	\end{minipage}
	
	\begin{minipage}[b]{.49\linewidth}
		\centering
		\centerline{\includegraphics[width=0.78\linewidth]{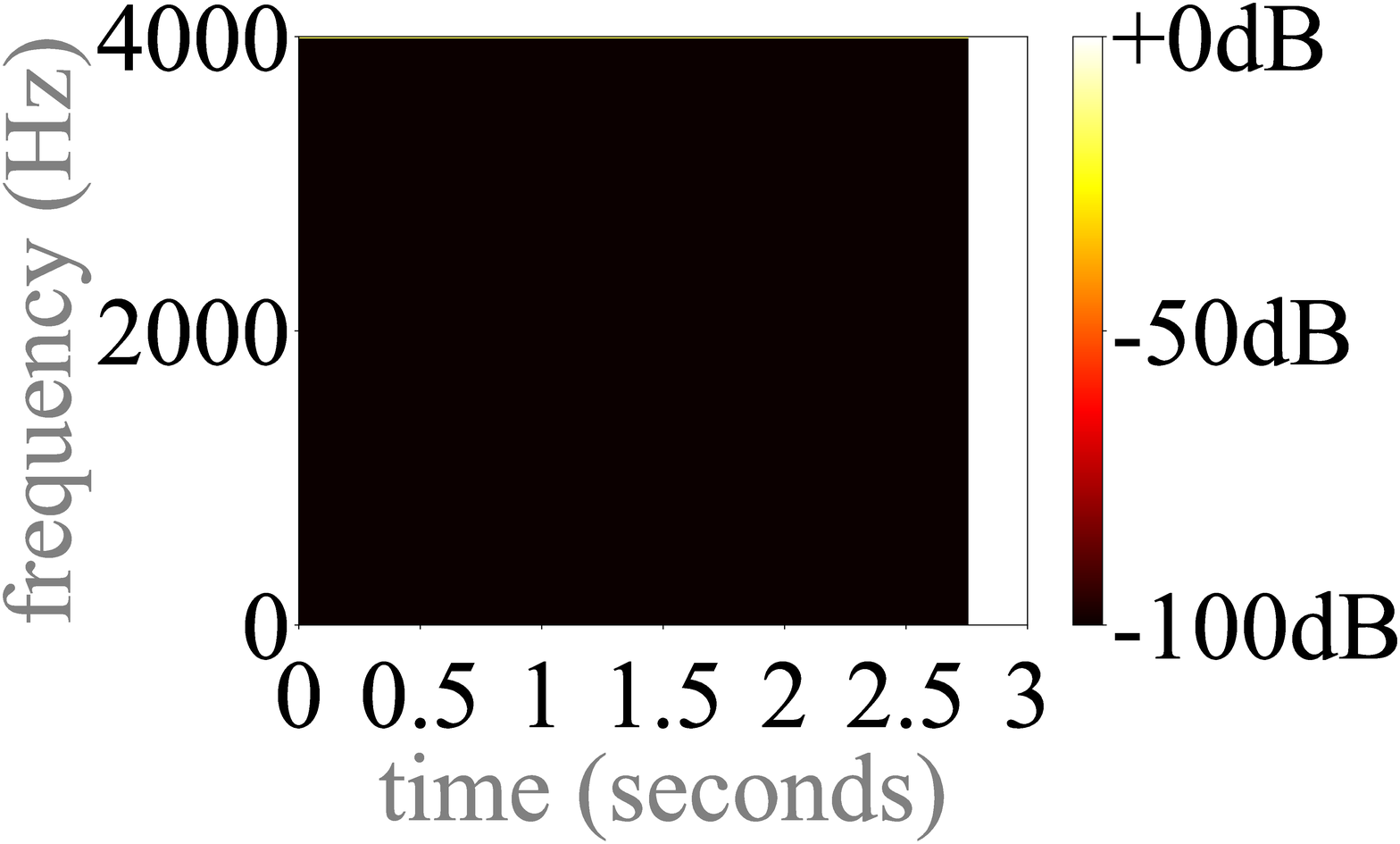}}
		\vspace{-0.3cm}
		{\mbox{(b) subpixel CNN: layer 1}}\medskip
	\end{minipage}
	\hfill 	\vspace{-0.4cm}
	\begin{minipage}[b]{0.49\linewidth}
		\centering
		\centerline{\includegraphics[width=0.78\linewidth]{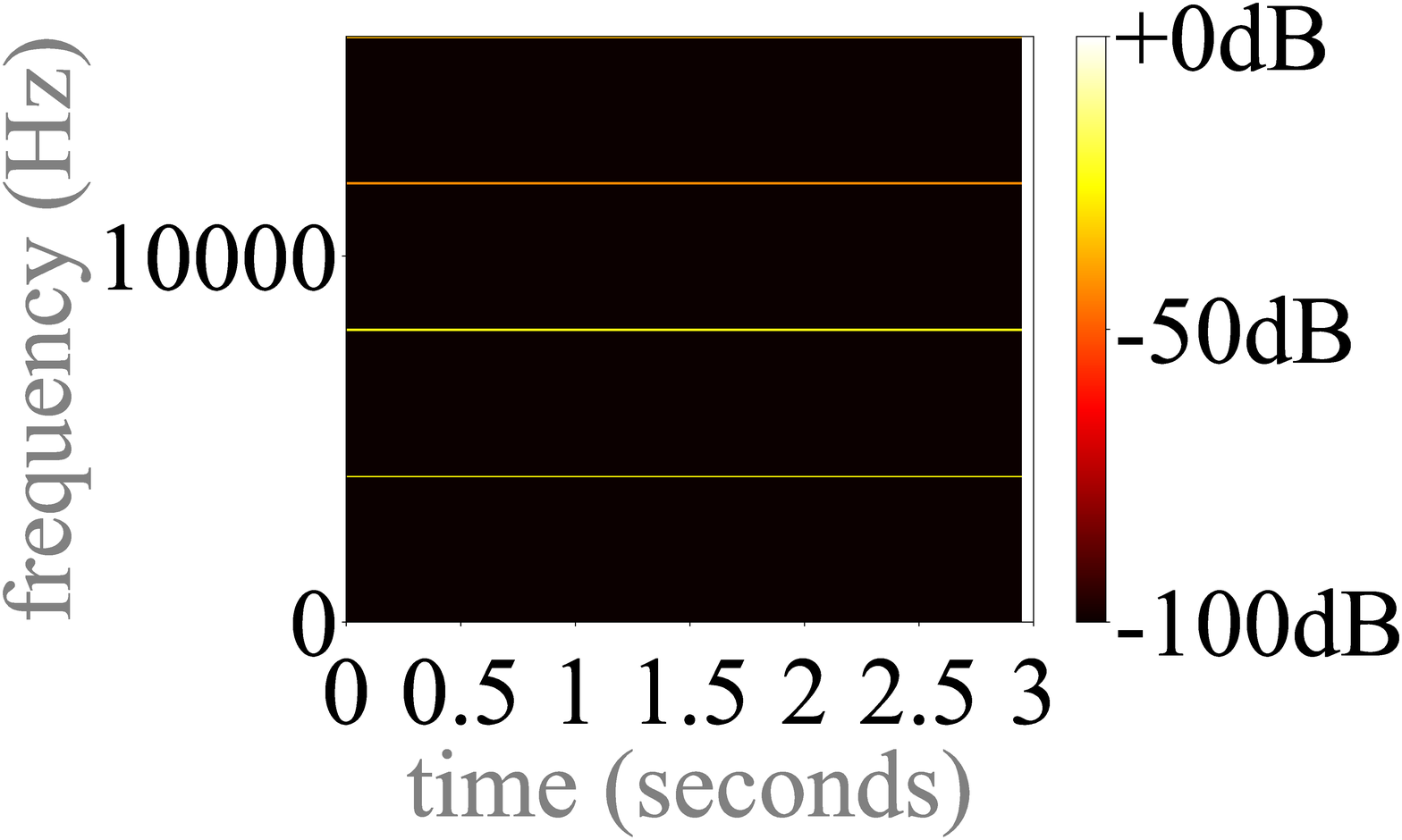}}
		\vspace{-0.3cm}
		{(e) subpixel CNN: layer 3}\medskip 
	\end{minipage}
	\vspace{-4mm}
	\caption{{\textbf{Subpixel CNN}: tonal artifacts after initialization. Each consecutive layer consists of a CNN (w/ filters of length 3 and stride of 1) + reshape via the periodic shuffle operation (upsample x2).}}
	\label{fig:subpixel CNN}
	\vspace{-5mm}
\end{figure}

\vspace{-2.5mm}
\section{\mbox{INTEPOLATION UPSAMPLERS}}
\vspace{-2mm}

{Interpolation + convolution} was proposed as an alternative to avoid transposed convolution artifacts~\cite{odena2016deconvolution}. 
It has been used, e.g., for music source separation~\cite{stoller2018wave}, speech enhancement~\cite{giri2019attention} and neural vocoding~\cite{gritsenko2020spectral}. 
While {interpolation} (e.g., linear or nearest neighbor) is effectively upsampling the signal, the subsequent convolution further transforms the upsampled signal with learnable weights.
Hence, the problematic transposed convolution is replaced by \mbox{interpolation + convolution} with the goal to avoid the above-described tonal artifacts (see Fig.~\ref{fig:interpolation} or Odena \textit{et al.}~\cite{odena2016deconvolution}).
Its main downsides are: (i) 
interpolation + convolution upsamplers require two layers instead of a single tranposed convolution layer, which increases the memory and computational footprint of the upsampler\footnote{Note that convolutions in interpolation upsamplers operate over longer (upsampled) signals, which is more expensive than operating over non-upsampled inputs as done by transposed or subpixel convolutions.}; and (ii) it can introduce filtering artifacts. 
Observe the filtering artifacts in Fig.~\ref{fig:interpolation} and \ref{fig:transposedcnn} (right): these de-emphasize the high-end frequencies of the spectrum. To understand filtering artifacts, note that interpolations can be implemented using convolutions---by first interpolating with zeros, an operator known as stretch~\cite{smith2007mathematics}, and later convolving with a pre-defined (non-learnable) filter. Linear interpolations can be implemented with triangular filters, a sinc$^2$($\cdot$) in the frequency domain; and nearest neighbor interpolation with rectangular filters, a sinc($\cdot$) in the frequency domain.
The side lobes of the linear interpolation filter, sinc$^2$($\cdot$), are lower than the nearest neighbor ones, sinc($\cdot$). For that reason, linear upsampling attenuates more the high-end frequencies than nearest neighbor upsampling (Fig.~\ref{fig:interpolation}). 
Unless using learnable interpolation filters~\cite{stoller2018wave}, interpolation filters cannot be fine-tuned and additional layers (like its subsequent learnable convolution) will have to compensate, if necessary, for the frequency response of the interpolation filter that introduces filtering artifacts. 
Hence, filtering artifacts color~\cite{pons2017timbre} the signal and are introduced by the {frequency response of the interpolation upsampler.}

	\vspace{-3mm}
\section{Subpixel Convolutions}
	\vspace{-2mm}

Based on {convolution + reshape}, subpixel CNN was proposed as an {efficient$^{1}$~upsampling layer~\cite{aitken2017checkerboard,shi2016real}.}
It has been used, e.g., for speech enhancement~\cite{pandey2020densely}, {bandwidth extension~\cite{kuleshov2017audio} and voice conversion~\cite{chou2018multi}.} 
The {convolution} upsamples the signal along the channel axis, and {reshape} is an operation called periodic shuffle~\cite{shi2016real} that reorders the convolution output to match the desired (upsampled) output shape. 
Subpixel CNNs~advantages are: (i)~it avoids overlap issues by construction, since {convolution + reshape} constrain it to~disallow overlapping; and~(ii) it~is computationally efficient because its convolutions operate over the original (non-upsampled) signal.$^1$
Its main drawback is that it can introduce tonal artifacts via the periodic shuffle operator (see Fig.~\ref{fig:subpixel CNN} or Aitken \textit{et al.}~\cite{aitken2017checkerboard}). Tonal artifacts emerge because
it upsamples consecutive samples based on convolutional filters having different weights, which can cause periodic patterns~\cite{aitken2017checkerboard}. 
\mbox{Aitken \textit{et al.}~\cite{aitken2017checkerboard}} proposed addressing these artifacts with an alternative initialization. 
However, nothing prevents these weights to degenerate into a solution that produces artifacts \mbox{again---i.e.,} tonal artifacts can emerge during and after training.

	\vspace{-3.2mm}

\section{Spectral replicas} 

	\vspace{-2mm}

\begin{figure}[tb]
		\vspace{-14mm}

	\begin{minipage}[b]{.49\linewidth}
		\centering
		\centerline{\includegraphics[width=0.78\linewidth]{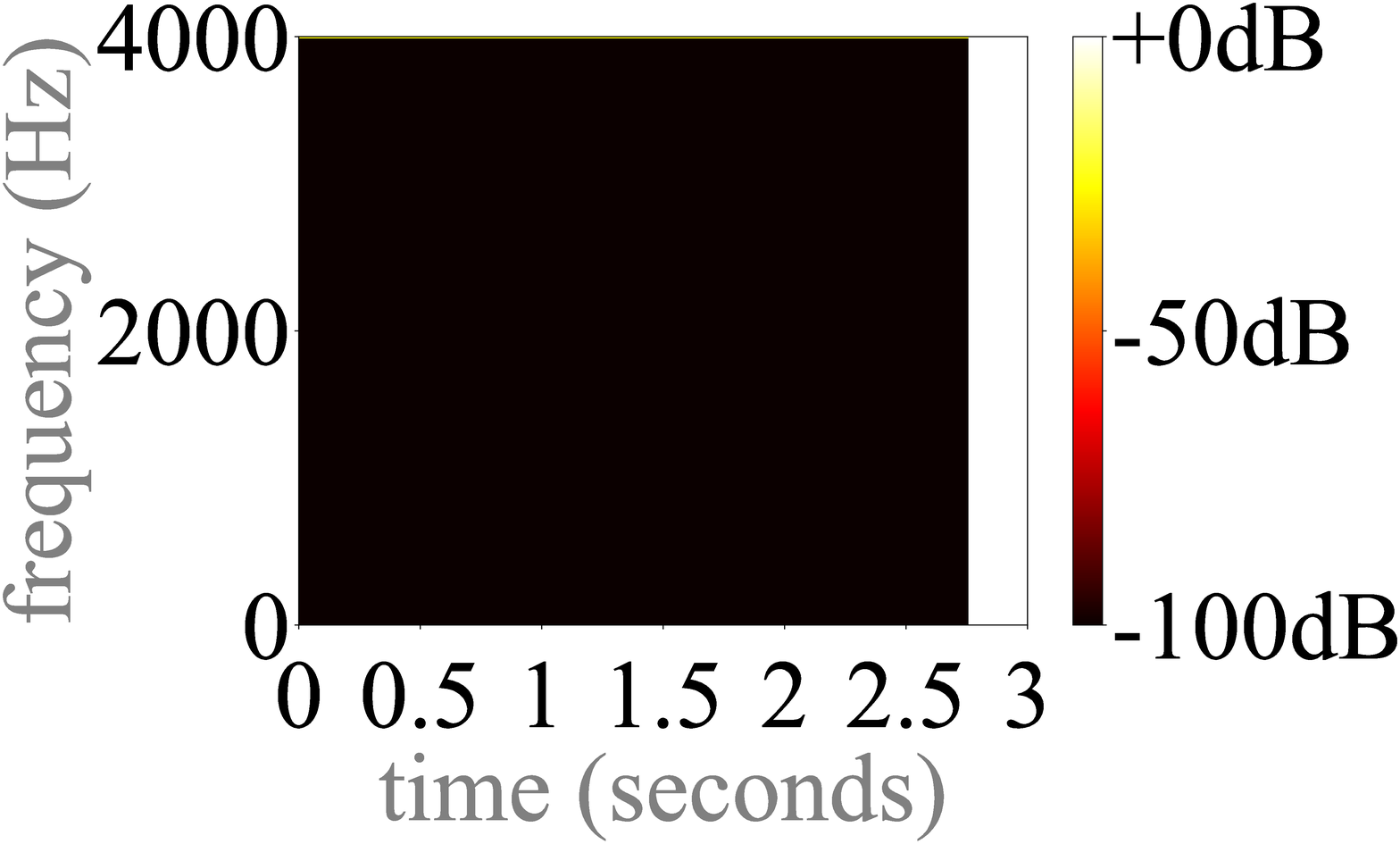}}
		\vspace{-0.32cm}
		{(a) Transposed CNN: layer 1}\medskip
	\end{minipage}
	\hfill		\vspace{-0.4cm}
	\begin{minipage}[b]{0.49\linewidth}
		\centering
		\centerline{\includegraphics[width=0.78\linewidth]{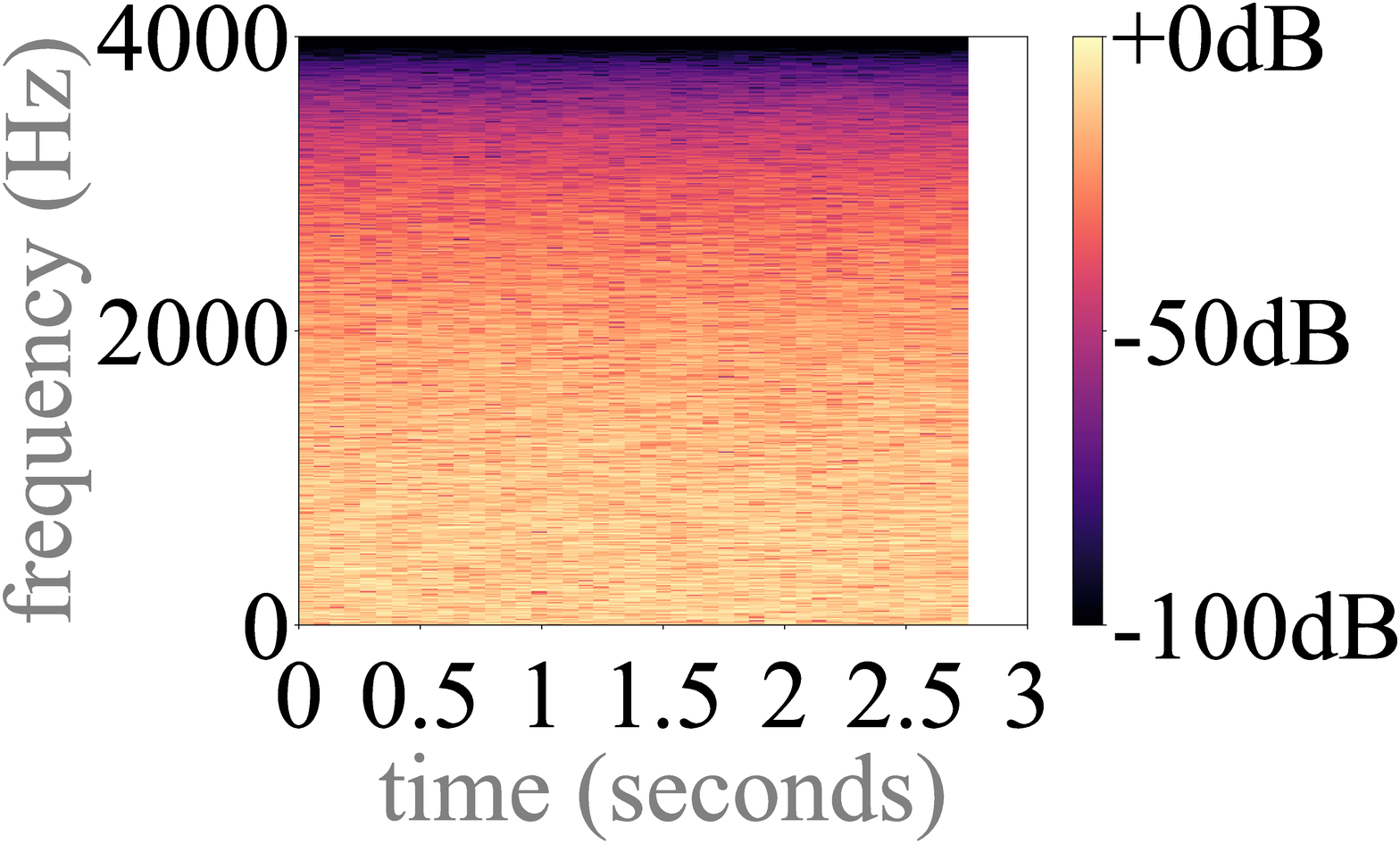}}
		\vspace{-0.32cm}
		{(d) Linear: layer 1}\medskip
	\end{minipage}
	\begin{minipage}[b]{.49\linewidth}
		\centering
		\centerline{\includegraphics[width=0.78\linewidth]{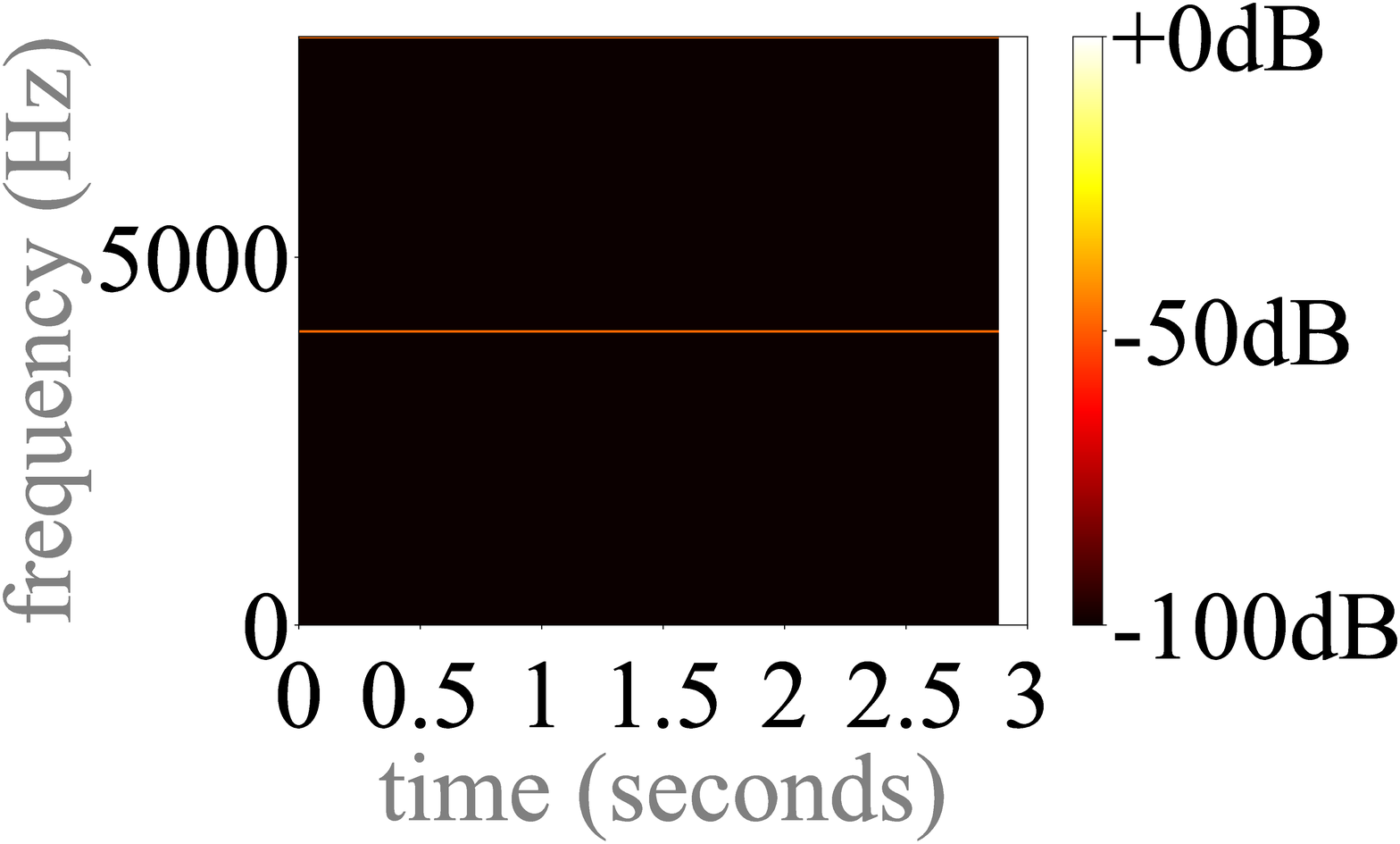}}
		\vspace{-0.32cm}
		{(b) Transposed CNN: layer 2}\medskip

	\end{minipage}
	\hfill 	\vspace{-0.4cm}
	\begin{minipage}[b]{0.49\linewidth}
		\centering
		\centerline{\includegraphics[width=0.78\linewidth]{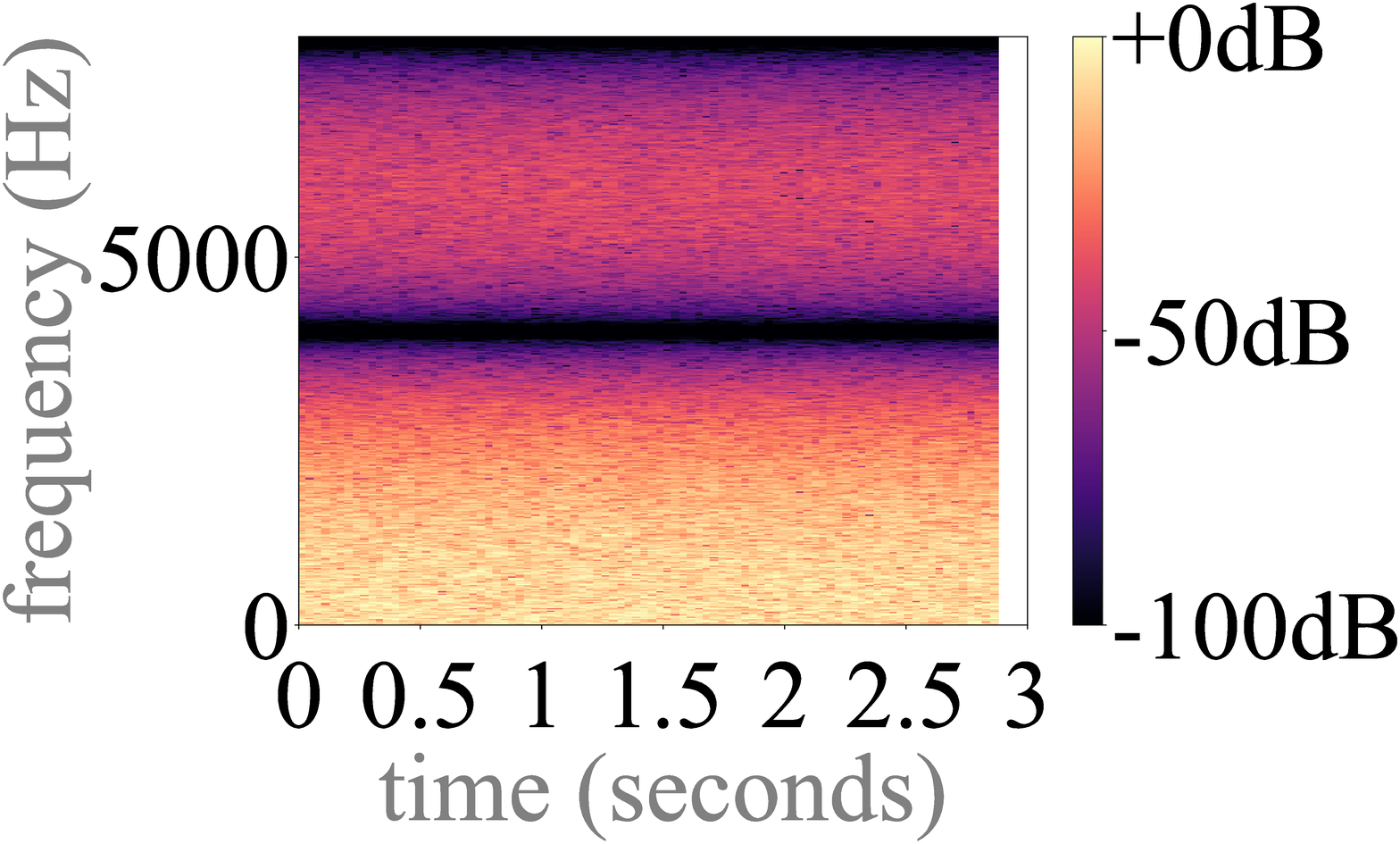}}
		\vspace{-0.32cm}
		{(e) Linear: layer 2}\medskip

	\end{minipage}
	\begin{minipage}[b]{.49\linewidth}
		\centering
		\centerline{\includegraphics[width=0.78\linewidth]{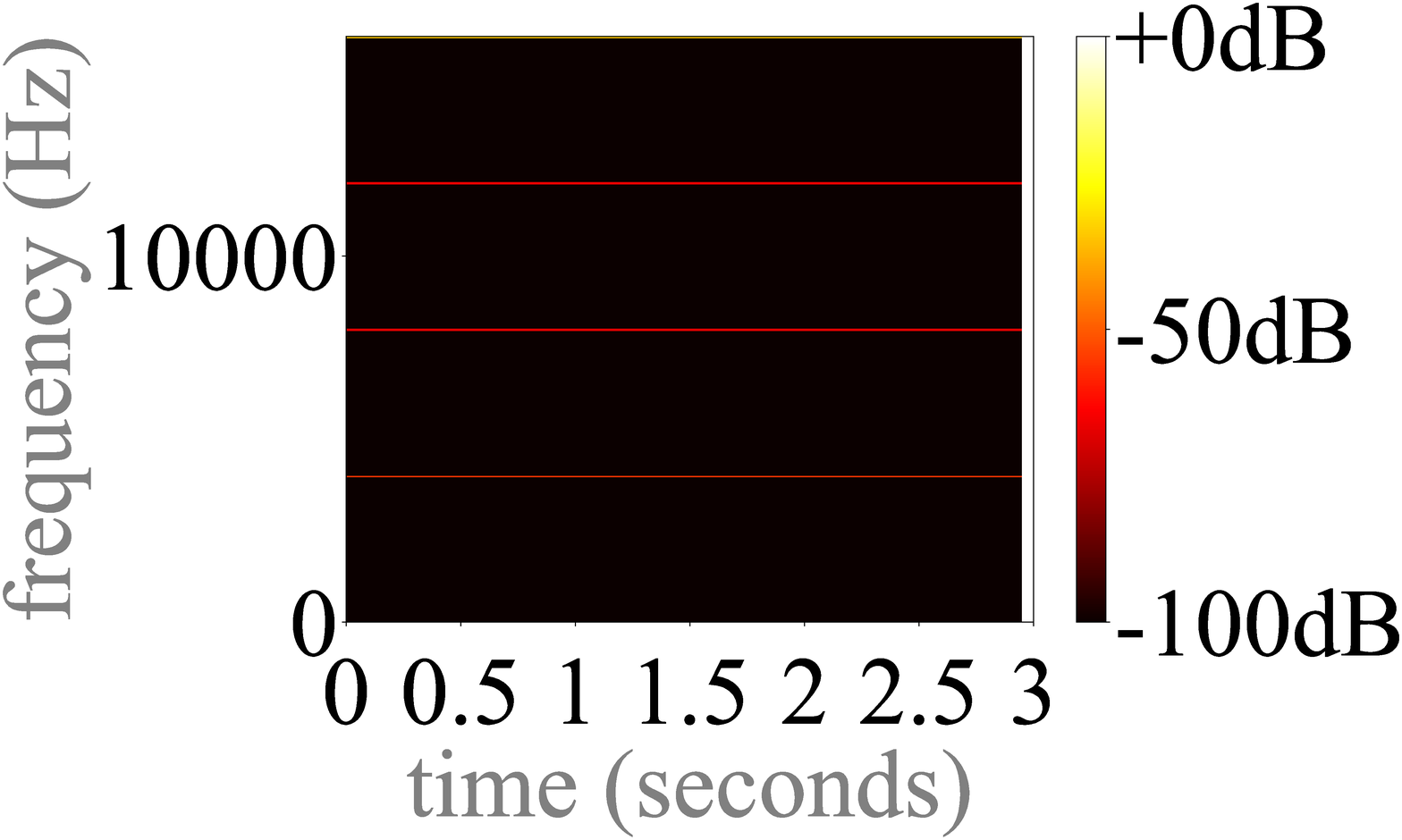}}
		\vspace{-0.32cm}
		{(c) Transposed CNN: layer 3}\medskip

	\end{minipage}
	\hfill 	\vspace{-0.4cm}
	\begin{minipage}[b]{0.49\linewidth}
		\centering
		\centerline{\includegraphics[width=0.78\linewidth]{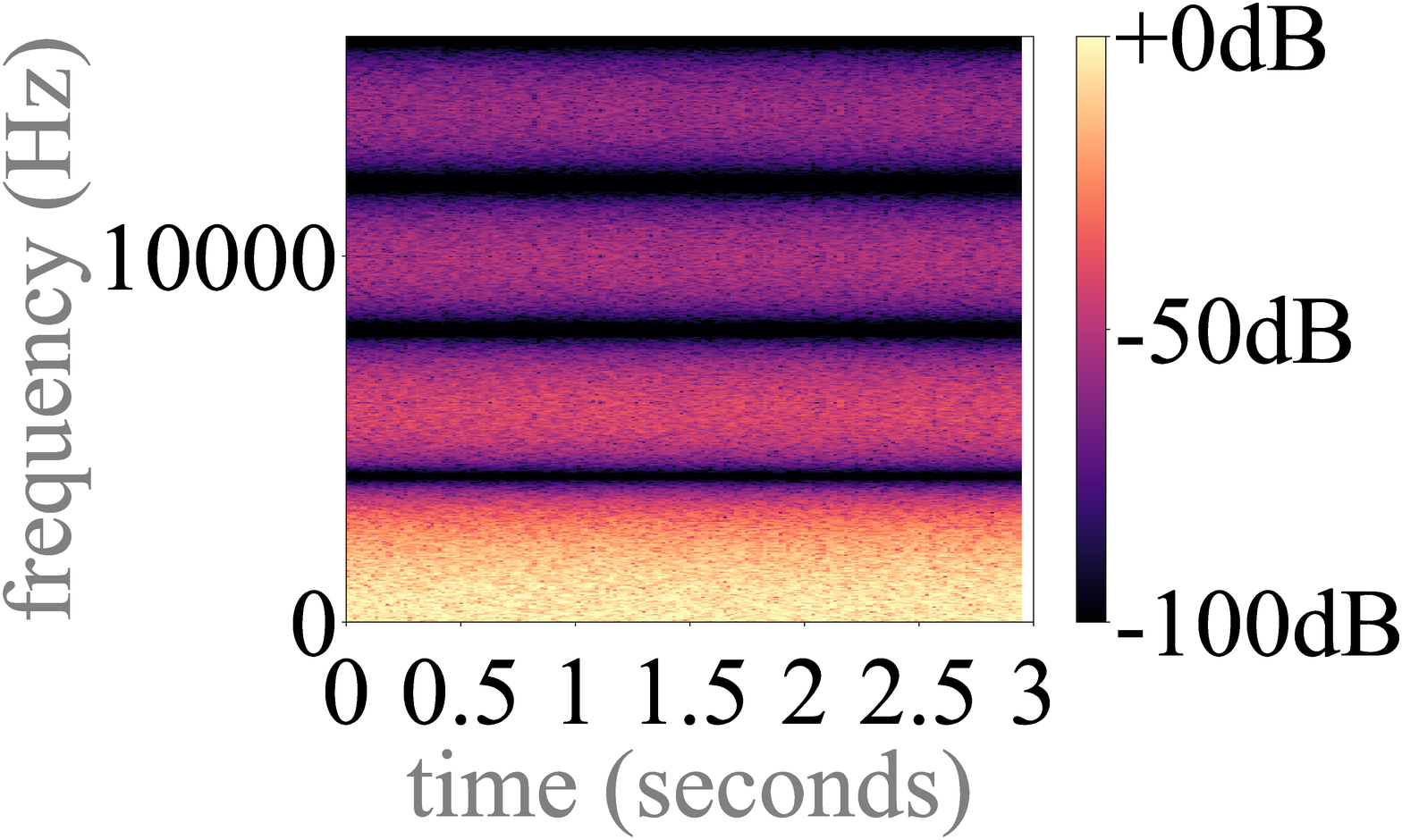}}
		\vspace{-0.32cm}
		{(f) Linear: layer 3}\medskip

	\end{minipage}
	\vspace{-5mm}
	
	\caption{{\textbf{Transposed CNN, linear interpolation}}: tonal and filtering artifacts, after initialization. Transposed convolution layers: \textit{length=4}, \textit{stride=2}. Linear interpolation layers: without the interleaved convolutions. Inputs at 4kHz: ones (left), white noise (right).}
	\label{fig:transposedcnn}
	\vspace{-2mm}
\end{figure}

Figs.~\ref{fig:interpolation},~\ref{fig:subpixel CNN} and~\ref{fig:transposedcnn} are illustrative because several artifacts interact: {(i)~tonal and filtering artifacts} introduced by problematic upsampling operations; and {(ii) spectral replicas} due to the bandwidth extension performed by each upsampling layer. 
From signal processing, we know that {spectral replicas} appear when discretizing a signal. 
Accordingly, when upsampling discrete signals one has to be vigilant of spectral replicas. 
Given that upsampling layers are effectively performing bandwidth extension, spectral replicas emerge while upsampling (see Fig.~\ref{fig:interpolation}, left).
Importantly, spectral replicas introduced by deeper layers (e.g.,~layers~2~{\footnotesize\&}~3) also include replicas of the artifacts introduced by previous layers \mbox{(e.g.,~layer 1 in  Figs.~\ref{fig:subpixel CNN} and~\ref{fig:transposedcnn}):}

\noindent \hspace{2mm}-- \textbf{Spectral replicas of tonal artifacts}. Upsampling tonal artifacts are introduced at a frequency of \mbox{``\textit{sampling rate / upsampling factor}",} the sampling rate being the one of the upsampled signal. For example: layer 1 outputs in Figs.~\ref{fig:subpixel CNN} and~\ref{fig:transposedcnn} (left) are at a sampling rate of~8kHz, because the 4kHz input was upsampled x2.\mbox{ Accordingly,} these upsampling layers introduce a tone at~4kHz. \mbox{When upsampling} with upcoming layers, the spectral replicas of~previously introduced tones are exposed---plus the employed upsampler introduces new tones.  In Figs.~\ref{fig:subpixel CNN} and~\ref{fig:transposedcnn} (left), the spectral replicas (at 8, 12, 16~kHz) interact with the tones introduced by each layer \mbox{(at 4, 8, 16 kHz).}

\noindent \hspace{2mm}-- \textbf{Spectral replicas of filtering artifacts}. Similarly, filtering artifacts are also replicated when upsampling---see Figs.~\ref{fig:all} (c), ~\ref{fig:interpolation} (right), \ref{fig:transposedcnn} (right).
This phenomenon is clearer in Fig.~\ref{fig:transposedcnn} (right) because the interleaved convolutions in Fig.~\ref{fig:interpolation} \mbox{(right) further color the spectrum.
}

\begin{figure}[th]
	\vspace{-2mm}
	\begin{minipage}[b]{.49\linewidth}
		\centering
		\centerline{\includegraphics[width=0.78\linewidth]{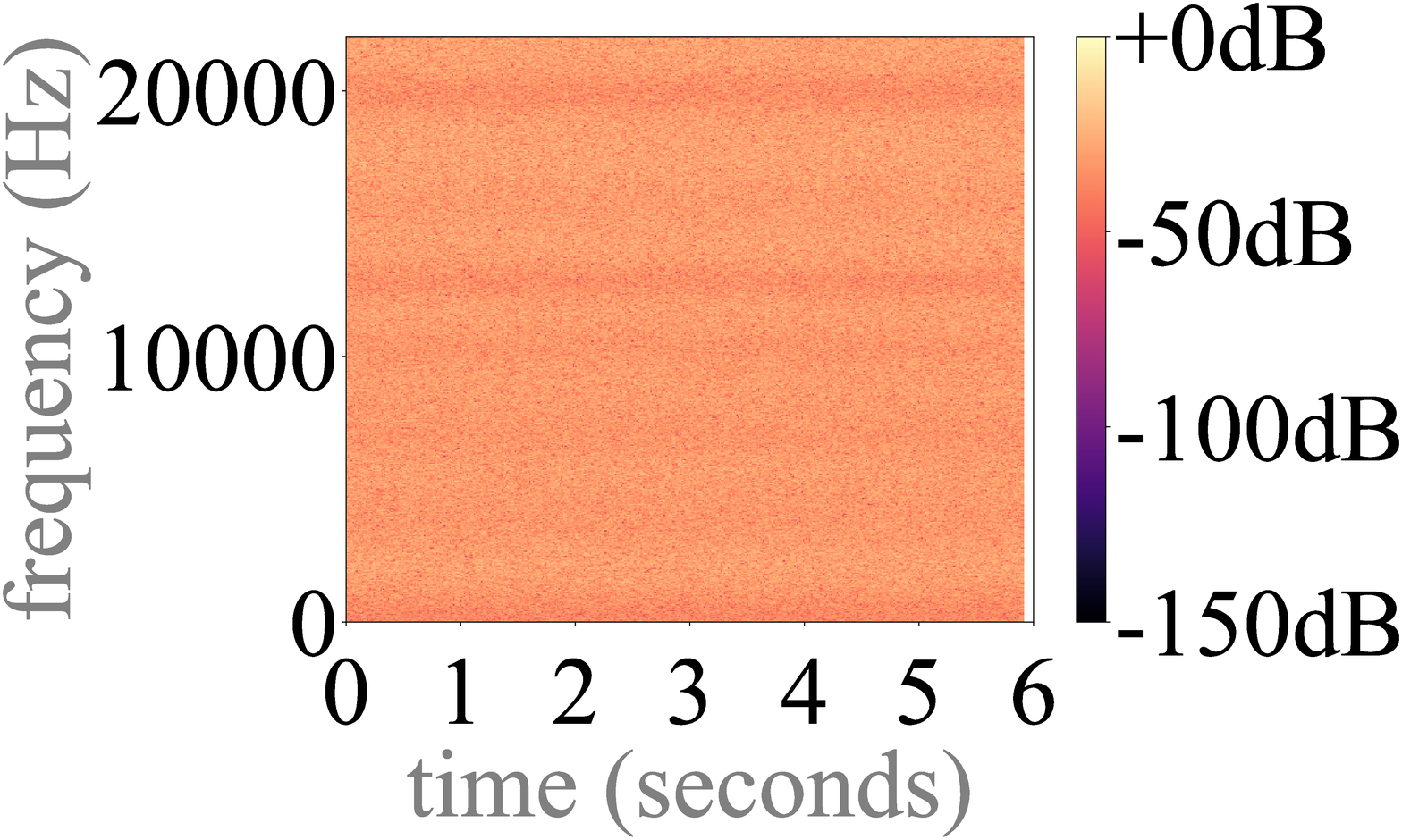}}
		\vspace{-0.3cm}
		{(a) Demucs (original) modification}\medskip

	\end{minipage}
	\hfill		\vspace{-0.5cm}
	\begin{minipage}[b]{0.49\linewidth}
		\centering
		\centerline{\includegraphics[width=0.78\linewidth]{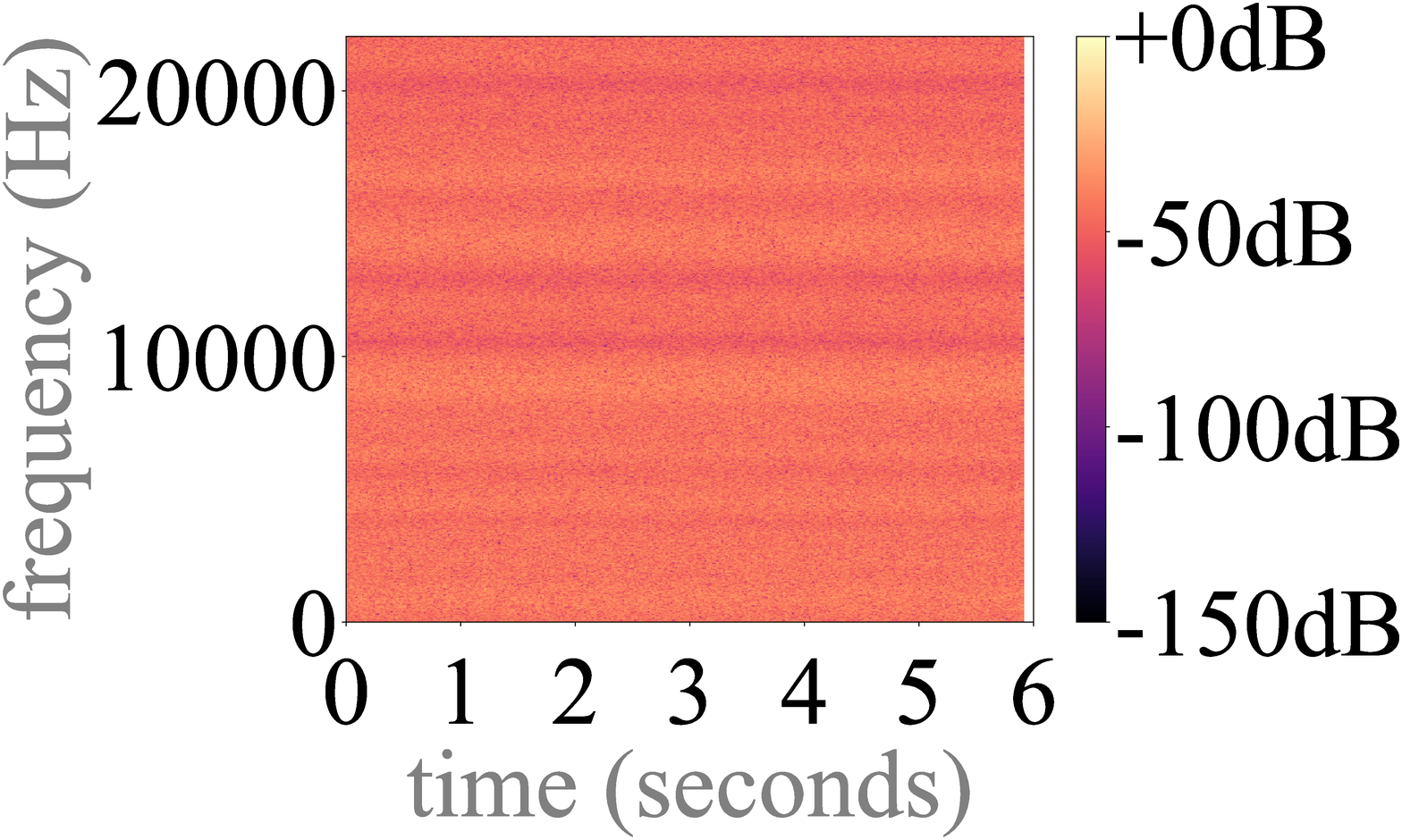}}
		\vspace{-0.3cm}
		{(d) Demucs (subpixel CNN) modification}\medskip

	\end{minipage}
	\vspace{-4mm}
	\caption{\textbf{Demucs modification after initialization}: no ReLUs in the first layer and no biases, to avoid spectral replicas of signal offsets.}
	\label{fig:modification}
	\vspace{-6mm}
\end{figure}

\noindent \hspace{2mm}-- \textbf{Spectral replicas of signal offsets}. Deep neural networks can include bias terms and ReLU non-linearities, which might introduce an offset to the resulting feature maps. Offsets are constant signals with zero frequency. Hence, its frequency transform contains an energy component at frequency zero. When upsampling, zero-frequency components are replicated in-band, introducing audible tonal artifacts. 
These signal offset replicas, however, can be removed with smart architecture designs. For example, via using the filtering artifacts (introduced by interpolation upsamplers) to attenuate the spectral replicas of signal offsets. Fig.~\ref{fig:transposedcnn} (right) shows that linear (but also nearest neigbor) upsamplers  attenuate such problematic frequencies, around ``\textit{sampling rate / upsampling factor}" where those tones appear.
Further, minor modifications to Demucs (just removing the ReLUs of the first layer\footnote{Due to the skip connections in Demucs, the first layer significantly affects the output. Also, removing all ReLUs could negatively affect training.} and the biases of the model) can also decrease the tonal artifacts after initialization (Fig.~\ref{fig:all} vs. Fig.~\ref{fig:modification}). While the modified Demucs architectures can still introduce tonal artifacts, via the problematic upsamplers that are used, the energy of the remaining tones is much less when compared to the tones introduced by the spectral replicas of signal offsets \mbox{(Fig.~\ref{fig:all} vs. Fig.~\ref{fig:modification})}. Note that mild tonal artifacts could be perceptually masked, since these are hardly noticeable under the presence of wide-band noise \mbox{(Fig.~\ref{fig:modification})}.

\noindent In signal processing, spectral replicas are normally removed with low-pass filters~\cite{smith2007mathematics}. Yet, upsampling layers are oftentimes stacked without those---so that upsamplers process the spectral replicas generated by previous upsampling layers~\cite{stoller2018wave,defossez2019music}. Note the effect of stacking upsamplers in Fig.~\ref{fig:interpolation}: the model colors~\cite{pons2017timbre} the spectral replicas from previous layers. 
For wide-band synthesis, however, it seems natural to allow that high-frequencies are available along the model.

	\vspace{-2.3mm}
\section{The role of training}
\label{sec:signals}
	\vspace{-2mm}

So far, we discussed that upsampling artifacts emerge because upsampling operators with problematic initializations are stacked one on top of another. However, nothing prevents the model from learning to compensate such artifacts.
While some did not use, e.g., transposed CNNs to avoid tonal artifacts~\cite{stoller2018wave,giri2019attention}---others aim at learning to correct these problematic initializations via training~\cite{defossez2019music,pandey2020densely}. 
We found that most speech upsamplers are based on transposed CNNs, with just a few exceptions~\cite{pandey2020densely,gritsenko2020spectral, giri2019attention}.
Only from the speech literature, we were unable to assess the impact of training one upsampler or another. Yet, music source separation works provide additional insights. 
WaveUnets (autoencoders based on linear upsamplers) are widely used, but their performance is poor compared to state-of-the-art models: $\approx$3 vs.~$\approx$5 dB SDR (signal-to-distortion ratio), see~Table~1. However, Demucs (a modified waveUnet relying on transposed convolutions) achieved competitive results: $\approx$5 dB SDR.~According to the literature, then, it seems that transposed CNNs are preferrable: these are widely used and achieve competitive results.

Here, we further study the role of learning when training neural upsamplers under comparable conditions.
We study Demucs-like models with 6 encoding blocks (with strided CNN, ReLU, GLU) and 6 decoding blocks (with GLU, full overlap transposed CNN, ReLU), connected via skip connections~\cite{defossez2019music}, with two LSTMs in the bottleneck (3200 units each). Strided and transposed convolution layers have 100, 200, 400, 800, 1600, 3200 filters of \textit{length=8} and \textit{stride=4}, respectively~\cite{defossez2019music}.
For our experiments, we change the transposed convolutions in Demucs for the different upsampling layers listed in Table 1 (top). Like the original Demucs, we use: very large models, of $\approx$700M parameters~\cite{defossez2019music}; weight rescaling, so that input and output signals are of the same magnitude after initialization~\cite{defossez2019music}; and their data augmentation schema, creating new mixes on-the-fly~\cite{defossez2019music,uhlich2017improving}.
We also use the MUSDB~\cite{musdb18} benchmark, that is composed of  stereo songs at 44.1 kHz. For each song, four stereo sources are extracted: vocals, bass, drums, other. In~Table~1 we report the average signal-to-distortion ratio (SDR)  over all sources~\cite{vincent2006performance}.\footnote{Following previous works~\cite{stoller2018wave,defossez2019music}: for every source, we report the median over all tracks of the median SDR over each test track of MUSDB~\cite{musdb18}.}
We optimize the L1 loss with Adam~\cite{kingma2014adam} for 600 epochs at a learning rate of 0.0002 {(0.0001 for the modified subpixel CNN)} with x4 v100 GPUs, using a batch size of 32. 
Despite their architectural issues and poor initialization, transposed and subpixel CNNs achieve the best SDR scores \mbox{(see Table 1). }
However, differently from what the literature conveys, interpolation-based models follow closely---with nearest neighbor upsamplers obtaining the best results.
The proposed modifications (without strong tones after initialization, see Fig.~\ref{fig:modification}) perform similarly to their poorly initialized counterparts. These results unveil the role of training: it helps overcoming the noisy initializations, caused by the problematic upsampling operators, to get state-of-the-art results.
Informal listening, however, reveals that tonal artifacts can emerge even after training, especially in silent parts and with out-of-distribution data (e.g., with sounds and conditions not seen during training). We find that nearest neighbor and linear interpolation models do not have this disadvantage, {although they achieve worse SDR scores.}

\begin{table}[t]
	\vspace{-11mm}
	\resizebox{\columnwidth}{!}{\begin{tabular}{l | c c c}
		
		\textbf{Music source separation (MUSDB~\cite{musdb18} benchmark)} & \textbf{SDR} $\uparrow$ & \textbf{epoch} & \textbf{\#parm} \\
		\hline	\hline
		Demucs-like (Fig.~\ref{fig:all}, b): transposed CNN (full-overlap) &  5.35& 319 s & 703M \\ 
		
		Demucs-like (Fig.~\ref{fig:all}, c): nearest neighbor interpolation & 5.17 & 423 s & 716M \\ 
				
		Demucs-like: linear interpolation & 4.62 & 430 s & 716M \\ 
				
		Demucs-like (Fig.~\ref{fig:all}, d): subpixel CNN & 5.38 & 311 s & 729M \\  
			\hline
		Modified (Fig.~\ref{fig:modification}, a): transposed CNN (full-overlap) &  5.37 & 326 s & 703M \\ 
		
		Modified (Fig.~\ref{fig:modification}, b): subpixel CNN & 5.38 & 315 s & 729M \\  
		
		\hline
		\hline
		WaveUnet~\cite{stoller2018wave}: linear intepolation & 3.23 & - &  10M  \\
		
		Demucs~\cite{defossez2019music}: transposed convolution (full-overlap) & \hspace{1.2mm}5.34\tablefootnote{With randomized equivariant stabilization: 5.58 dB SDR~\cite{defossez2019music}. We do not use this technique for simplicity, since it does not relate to the training phase.} & - &  648M   \\
		
		OpenUnmix~\cite{stoter19}: spectrogram-based & 5.36& - &  8.9M   \\
		
		Sams-net~\cite{samsnet}: spectrogram-based &  5.65 & - & 3.7M   \\

	\end{tabular}}
	\vspace{-4mm}
	\caption{Demucs-like models with different upsamplers (top), against the {state-of-the-art (bottom). SDR (dB): the higher the better.}}
	\label{tab:table-name}
	\vspace{-6mm}
\end{table}

	\vspace{-3mm}
\section{SUMMARY \& REMARKS}
		\vspace{-2mm}

Upsamplers are a key element for developing computationally efficient and high-fidelity neural audio synthesizers. 
Given their importance, together with the fact that the audio literature only provides sparse and unorganized insights~\cite{donahue2018adversarial,stoller2018wave,kumar2019melgan,pandey2020densely}, our work is aimed at advancing and consolidating our current understanding of neural upsamplers.
We discussed several sources of tonal artifacts: some relate to the transposed convolution setup or weight initialization, others to architectural choices not related to transposed convolutions (e.g.,~with subpixel CNNs), and others relate to the way learning is defined (e.g.,~with adversarial or deep feature losses).
While several works assume to resolve the tonal artifacts via learning from data, others looked at alternative possibilities which, by construction, omit tonal artifacts---these include: interpolation upsamplers, that can introduce filtering artifacts.
Further, upsampling artifacts can be emphasized by deeper layers via exposing their spectral replicas.
We want to remark that any transposed convolution setup, even with full or no overlap, produces a poor initialization due to {the weights initialization issue}. Further,  subpixel CNNs can also introduce tonal artifacts. In both cases, training is responsible to compensate for any upsampling artifact.
Finally, the interpolation upsamplers we study do not introduce tonal artifacts, what is perceptually preferable, but they {achieve worse SDR results and can introduce filtering artifacts.}

\newpage

\section{APPENDIX: A NOTE ON SPECTROGRAMS}


We compute the above spectrograms (STFT: short-time fourier transform) with Librosa~\cite{mcfee2015librosa}, that features a \textit{center} flag in its STFT implementation.
If \textit{center=True}, each frame is centered around the given time-stamp and the signal is padded if necessary. As a result of this padding, STFTs include boundary artifacts---but the temporal alignment is precise. Note this behaviour in Fig. 10 (right), where boundary artifacts are introduced but signals are perfectly aligned in the temporal axis.
Alternatively, if \textit{center=False}, the signal is not padded and frames are aligned to the left. Note this behaviour in Fig. 10 (left) where no boundary artifacts are introduced---but signals are aligned to the left, and white stripes appear on the right. For our study we set \textit{center=False}, because we want to avoid mixing the boundary artifacts introduced by the STFT with potential boundary artifacts introduced by upsampling layers.

\begin{figure}[h]
\vspace{-3mm}
	
	\begin{minipage}[b]{.49\linewidth}
		\centering
		\centerline{\includegraphics[width=0.78\linewidth]{Fig8l_1}}
		\vspace{-0.32cm}
		{(a) Transposed CNN: layer 1}\medskip
	\end{minipage}
	\hfill		\vspace{-0.4cm}
	\begin{minipage}[b]{0.49\linewidth}
		\centering
		\centerline{\includegraphics[width=0.78\linewidth]{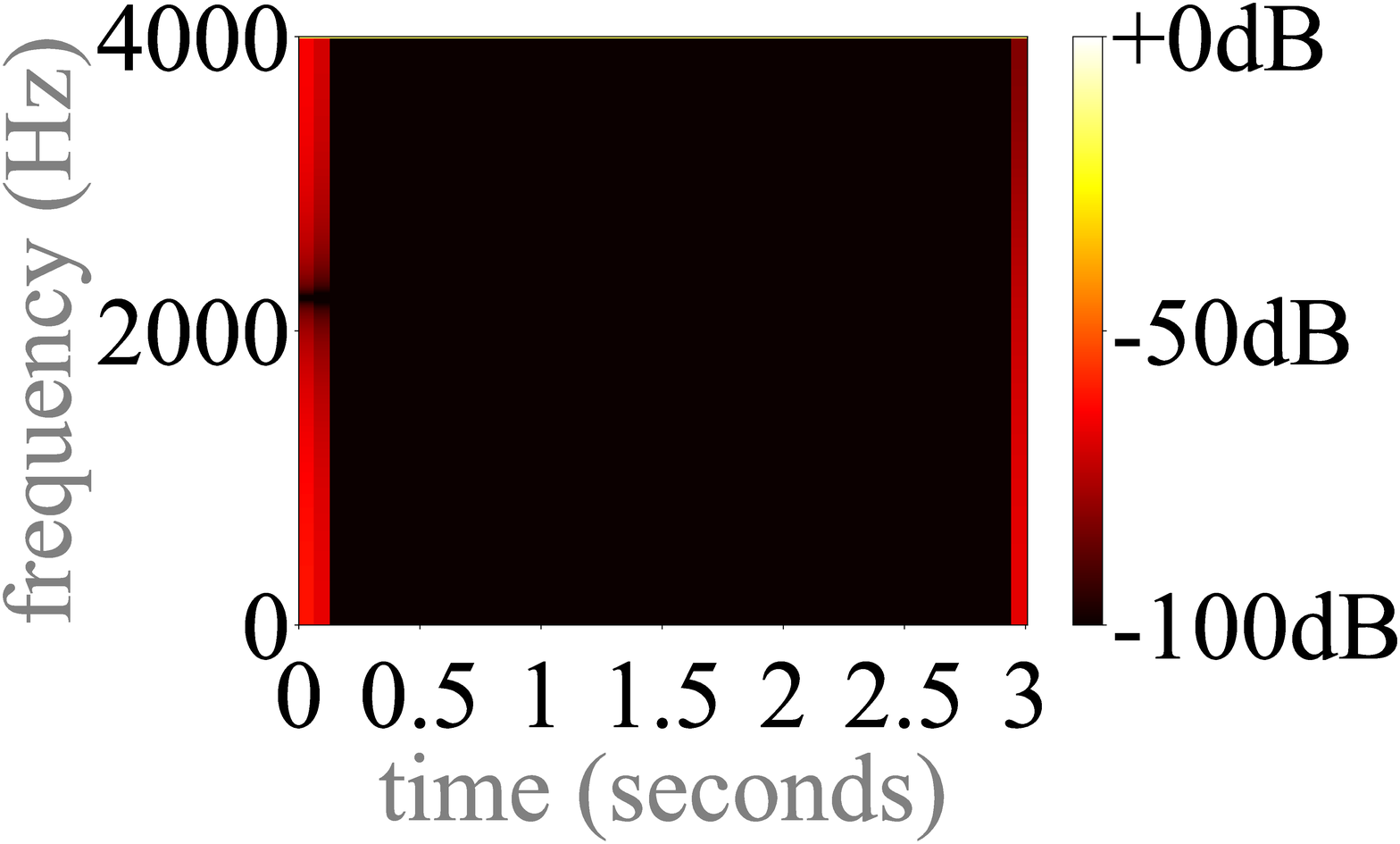}}
		\vspace{-0.32cm}
		{(d) Transposed CNN: layer 1}\medskip
	\end{minipage}
	\begin{minipage}[b]{.49\linewidth}
		\centering
		\centerline{\includegraphics[width=0.78\linewidth]{Fig8l_2}}
		\vspace{-0.32cm}
		{(b) Transposed CNN: layer 2}\medskip
		
	\end{minipage}
	\hfill 	\vspace{-0.4cm}
	\begin{minipage}[b]{0.49\linewidth}
		\centering
		\centerline{\includegraphics[width=0.78\linewidth]{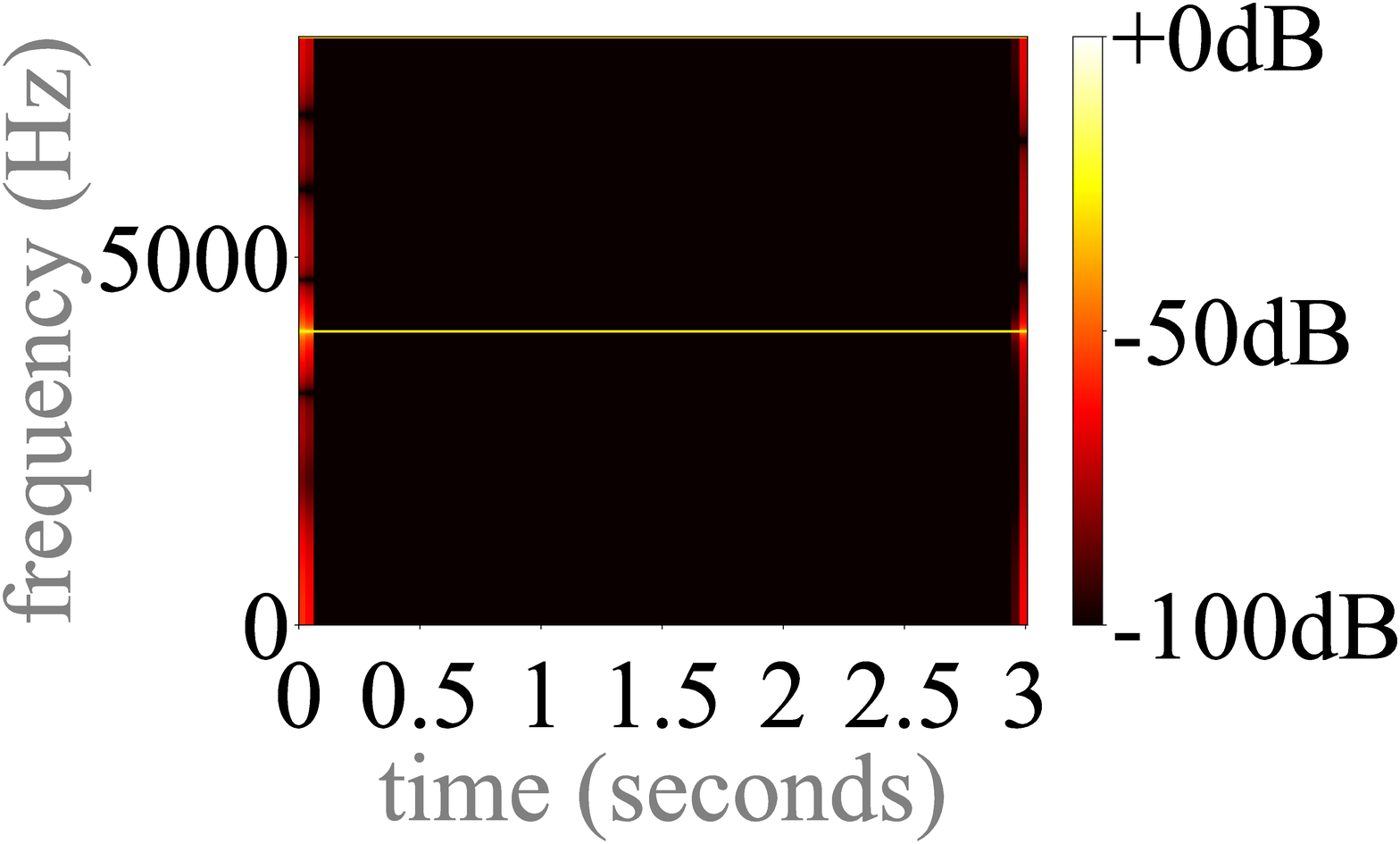}}
		\vspace{-0.32cm}
		{(e) Transposed CNN: layer 2}\medskip
		
	\end{minipage}
	\begin{minipage}[b]{.49\linewidth}
		\centering
		\centerline{\includegraphics[width=0.78\linewidth]{Fig8l_3}}
		\vspace{-0.32cm}
		{(c) Transposed CNN: layer 3}\medskip
		
	\end{minipage}
	\hfill 	\vspace{-0.4cm}
	\begin{minipage}[b]{0.49\linewidth}
		\centering
		\centerline{\includegraphics[width=0.78\linewidth]{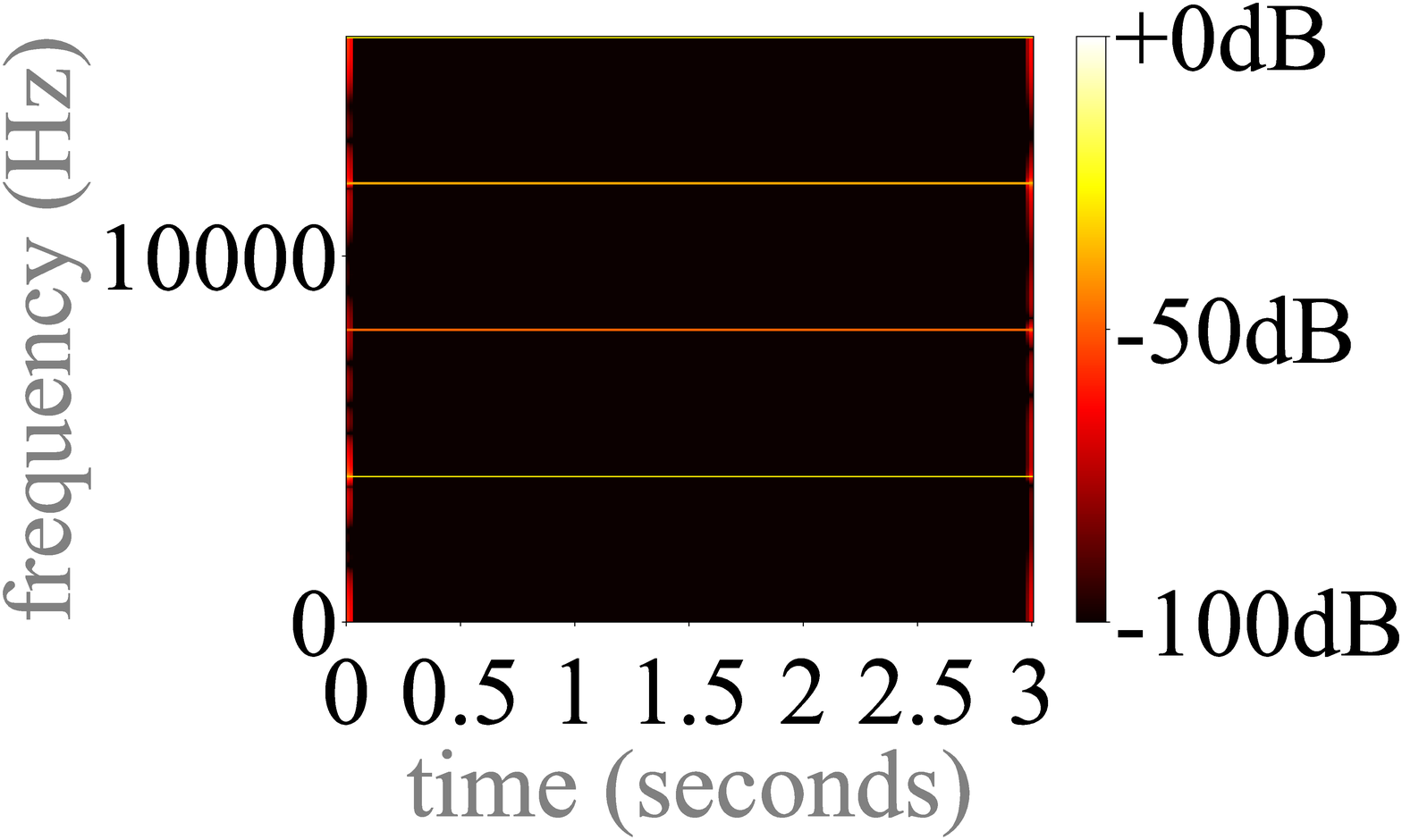}}
		\vspace{-0.32cm}
		{(f) Transposed CNN: layer 3}\medskip
		
	\end{minipage}
	\vspace{-3mm}
	
	\caption{Transposed CNNs (\textit{length=4}, \textit{stride=2}): tonal artifacts after initialization. Inputs at 4kHz: ones. As Fig. 8, with different STFT setups.
	\textbf{Left}: STFT (\textit{center=False}). \textbf{Right}: STFT (\textit{center=True}).}
\end{figure}

\noindent Finally, white stripes are wider in layers 1 and 2 because we use the same STFT setup (\textit{n\_fft=2048, hop\_length=512}) for all signals, even though each layer operates at a different sampling rate (8kHz, 16kHz and 32kHz respectively).

\bibliographystyle{IEEEbib}
\bibliography{strings,refs}

\begin{thebibliography}{10}

\bibitem{donahue2018adversarial}
Chris Donahue, Julian McAuley, and Miller Puckette,
\newblock ``Adversarial audio synthesis,''
\newblock in {\em ICLR}, 2019.

\bibitem{stoller2018wave}
Daniel Stoller, Sebastian Ewert, and Simon Dixon,
\newblock ``Wave-u-net: A multi-scale neural network for end-to-end audio
  source separation,''
\newblock {\em ISMIR}, 2018.

\bibitem{defossez2019music}
Alexandre D{\'e}fossez, Nicolas Usunier, L{\'e}on Bottou, and Francis Bach,
\newblock ``Music source separation in the waveform domain,''
\newblock {\em arXiv}, 2019.

\bibitem{kumar2019melgan}
Kundan Kumar, Rithesh Kumar, Thibault de~Boissiere, Lucas Gestin, Wei~Zhen
  Teoh, Jose Sotelo, Alexandre de~Br{\'e}bisson, Yoshua Bengio, and Aaron~C
  Courville,
\newblock ``Melgan: Generative adversarial networks for conditional waveform
  synthesis,''
\newblock in {\em NeurIPS}, 2019.

\bibitem{oord2016wavenet}
Aaron van~den Oord, Sander Dieleman, Heiga Zen, Karen Simonyan, Oriol Vinyals,
  Alex Graves, Nal Kalchbrenner, Andrew Senior, and Koray Kavukcuoglu,
\newblock ``Wavenet: A generative model for raw audio,''
\newblock {\em arXiv}, 2016.

\bibitem{rethage2018wavenet}
Dario Rethage, Jordi Pons, and Xavier Serra,
\newblock ``A wavenet for speech denoising,''
\newblock in {\em ICASSP}, 2018.

\bibitem{prenger2019waveglow}
Ryan Prenger, Rafael Valle, and Bryan Catanzaro,
\newblock ``Waveglow: A flow-based generative network for speech synthesis,''
\newblock in {\em ICASSP}, 2019.

\bibitem{serra2019blow}
Joan Serr{\`a}, Santiago Pascual, and Carlos~Segura Perales,
\newblock ``Blow: a single-scale hyperconditioned flow for non-parallel
  raw-audio voice conversion,''
\newblock in {\em NeurIPS}, 2019.

\bibitem{pascual2017segan}
Santiago Pascual, Antonio Bonafonte, and Joan Serr{\`a},
\newblock ``Segan: Speech enhancement generative adversarial network,''
\newblock {\em Interspeech}, 2017.

\bibitem{pandey2020densely}
Ashutosh Pandey and DeLiang Wang,
\newblock ``Densely connected neural network with dilated convolutions for
  real-time speech enhancement in the time domain,''
\newblock in {\em ICASSP}, 2020.

\bibitem{binkowski2019high}
Miko{\l}aj Bi{\'n}kowski, Jeff Donahue, Sander Dieleman, Aidan Clark, Erich
  Elsen, Norman Casagrande, Luis~C Cobo, and Karen Simonyan,
\newblock ``High fidelity speech synthesis with adversarial networks,''
\newblock in {\em ICLR}, 2019.

\bibitem{kuleshov2017audio}
Volodymyr Kuleshov, S~Zayd Enam, and Stefano Ermon,
\newblock ``Audio super resolution using neural networks,''
\newblock {\em arXiv}, 2017.

\bibitem{chou2018multi}
Ju-chieh Chou, Cheng-chieh Yeh, Hung-yi Lee, and Lin-shan Lee,
\newblock ``Multi-target voice conversion without parallel data by
  adversarially learning disentangled audio representations,''
\newblock {\em arXiv}, 2018.

\bibitem{gritsenko2020spectral}
Alexey~A Gritsenko, Tim Salimans, Rianne van~den Berg, Jasper Snoek, and Nal
  Kalchbrenner,
\newblock ``A spectral energy distance for parallel speech synthesis,''
\newblock {\em arXiv}, 2020.

\bibitem{odena2016deconvolution}
Augustus Odena, Vincent Dumoulin, and Chris Olah,
\newblock ``Deconvolution and checkerboard artifacts,''
\newblock {\em Distill}, vol. 1, no. 10, pp. e3, 2016.

\bibitem{germain2019speech}
Francois~G Germain, Qifeng Chen, and Vladlen Koltun,
\newblock ``Speech denoising with deep feature losses,''
\newblock {\em Interspeech}, 2019.

\bibitem{olah2017feature}
Chris Olah, Alexander Mordvintsev, and Ludwig Schubert,
\newblock ``Feature visualization,''
\newblock {\em Distill}, vol. 2, no. 11, pp. e7, 2017.

\bibitem{nguyen2016synthesizing}
Anh Nguyen, Alexey Dosovitskiy, Jason Yosinski, Thomas Brox, and Jeff Clune,
\newblock ``Synthesizing the preferred inputs for neurons in neural networks
  via deep generator networks,''
\newblock in {\em NeurIPS}, 2016.

\bibitem{mahendran2015understanding}
Aravindh Mahendran and Andrea Vedaldi,
\newblock ``Understanding deep image representations by inverting them,''
\newblock in {\em CVPR}, 2015.

\bibitem{mordvintsev2015inceptionism}
Alexander Mordvintsev, Christopher Olah, and Mike Tyka,
\newblock ``Inceptionism: Going deeper into neural networks,''
\newblock {\em Google Research Blog}, 2015.

\bibitem{aitken2017checkerboard}
Andrew Aitken, Christian Ledig, Lucas Theis, Jose Caballero, Zehan Wang, and
  Wenzhe Shi,
\newblock ``Checkerboard artifact free sub-pixel convolution: A note on
  sub-pixel convolution, resize convolution and convolution resize,''
\newblock {\em arXiv}, 2017.

\bibitem{giri2019attention}
Ritwik Giri, Umut Isik, and Arvindh Krishnaswamy,
\newblock ``Attention wave-u-net for speech enhancement,''
\newblock in {\em WASPAA}, 2019.

\bibitem{smith2007mathematics}
Julius~Orion Smith,
\newblock {\em Mathematics of the discrete Fourier transform (DFT): with audio
  applications},
\newblock Julius Smith, 2007.

\bibitem{pons2017timbre}
Jordi Pons, Olga Slizovskaia, Rong Gong, Emilia G{\'o}mez, and Xavier Serra,
\newblock ``Timbre analysis of music audio signals with convolutional neural
  networks,''
\newblock in {\em EUSIPCO}, 2017.

\bibitem{shi2016real}
Wenzhe Shi, Jose Caballero, Ferenc Husz{\'a}r, Johannes Totz, Andrew~P Aitken,
  Rob Bishop, Daniel Rueckert, and Zehan Wang,
\newblock ``Real-time single image and video super-resolution using an
  efficient sub-pixel convolutional neural network,''
\newblock in {\em CVPR}, 2016.

\bibitem{uhlich2017improving}
Stefan Uhlich, Marcello Porcu, Franck Giron, Michael Enenkl, Thomas Kemp, Naoya
  Takahashi, and Yuki Mitsufuji,
\newblock ``Improving music source separation based on deep neural networks
  through data augmentation and network blending,''
\newblock in {\em ICASSP}, 2017.

\bibitem{musdb18}
Zafar Rafii, Antoine Liutkus, Fabian-Robert St{\"o}ter, Stylianos~Ioannis
  Mimilakis, and Rachel Bittner,
\newblock ``The {MUSDB18} corpus for music separation,'' 2017.

\bibitem{vincent2006performance}
Emmanuel Vincent, R{\'e}mi Gribonval, and C{\'e}dric F{\'e}votte,
\newblock ``Performance measurement in blind audio source separation,''
\newblock {\em IEEE Transactions on Audio, Speech, and Language Processing},
  vol. 14, no. 4, pp. 1462--1469, 2006.

\bibitem{kingma2014adam}
Diederik~P Kingma and Jimmy Ba,
\newblock ``Adam: A method for stochastic optimization,''
\newblock {\em arXiv}, 2014.

\bibitem{stoter19}
Fabian-Robert St{\"o}ter, S.~Uhlich, A.~Liutkus, and Y.~Mitsufuji,
\newblock ``Open-unmix - a reference implementation for music source
  separation,''
\newblock {\em Journal of Open Source Software}, 2019.

\bibitem{samsnet}
Haowen Hou Ming~Li Tingle~Li, Jiawei~Chen,
\newblock ``Sams-net: A sliced attention-based neural network for music source
  separation,''
\newblock {\em arXiv}, 2019.

\bibitem{mcfee2015librosa}
Brian McFee, Colin Raffel, Dawen Liang, Daniel~PW Ellis, Matt McVicar, Eric
  Battenberg, and Oriol Nieto,
\newblock ``librosa: Audio and music signal analysis in python,''
\newblock in {\em Proceedings of the 14th python in science conference}.
  Citeseer, 2015, vol.~8, pp. 18--25.

\end{thebibliography}

\end{document}